# Conical Intersections and Charge Localization in a Minimal Model of a Monomethine Dye System: Degenerate Case

*Seth Olsen and Ross H. McKenzie*

Centre for Organic Photonics & Electronics and School of Mathematics and Physics, The University of Queensland, St. Lucia, QLD 4072 Australia

s.olsen1@uq.edu.au

TITLE RUNNING HEAD (Word Style "AF_Title_Running_Head"). If you are submitting your paper to a journal that requires a title running head (condensed title appearing at the top of the journal page), please provide a 50-character or less summary of the title of the manuscript.

CORRESPONDING AUTHOR FOOTNOTE (Word Style "FA_Corresponding_Author_Footnote"). Clarify all corresponding authors' addresses by accompanying footnotes if they are not apparent from the affiliation line. Telephone numbers, fax numbers, and e-mail addresses may all be included in the corresponding author footnotes. If an author's address is different than the one given in the affiliation line, this information may also be included here.

ABSTRACT   We propose a simple valence-bond-based model Hamiltonian to describe the photoisomerization of methine-bridged systems. These include a broad class of large conjugated organic molecules including diarylmethane and cyanine dyes, and the chromophores of fluorescent proteins. The Hamiltonian is defined over an electronic state space spanned by three valence-bond structures, which are themselves defined over three fragment π orbitals defining a single methine moiety. Analogous valence-bond structures can be drawn for cases where the three orbitals contain four or two electrons, describing systems where formal charge – positive or negative, respectively – resonates amongst the fragments in tandem with a single π bond. We parameterize the model using three independent parameters. The parameters include two angles representing torsion between the ends and the methine bridge, and a positive real number that describes a ratio between end-bridge and end-end coupling. The third parameter can be geometrically interpreted as a combination of symmetric stretching and a bridge internal angle bend. The model fulfills the most basic expectations for a double bond photoisomerization model: 1) it describes favorable twisting motions in the excited state and 2) it describes twisting leads to rapid non-radiative decay via conical intersections. We examine in detail the topology, topography and extent of the conical intersection seams, which can be visualized in their entirety over the three-dimensional parameter space. Important properties of the model include 1) two regimes of the coupling ratio where the first excited state intersects the ground and a higher excited state along orthogonal twisting coordinates or along the same coordinate 2) first and higher excited state intersections occurring in regions of little or no twisting and 3) charge localization connected with excited-state twisting which changes discontinuously across a conical intersection. The model is promising as a simple extension of empirical force fields in order to describe photoisomerization processes in methine-bridged systems and as a starting point for more elaborate models which may include antisymmetric internal couplings or which may be interfaced with environmental or quantum dynamical models.

BRIEFS (WORD Style "BH_Briefs"). If you are submitting your paper to a journal that requires a brief, provide a one-sentence synopsis for inclusion in the Table of Contents.

## INTRODUCTION

The efficient description of photochemical excited states has been[1,2] and continues to be[3] a challenge to theory. However, meeting this challenge is a prerequisite to the understanding and simulation of condensed phase photochemical processes, where the electronic structure problem is only one aspect of the challenge[4]. Such progress is essential to our understanding of crucial biophysical processes[5] and to the development of biomimetic alternate energy technologies[6]. Although great strides have been made in the *ab initio* determination of excited states of small to medium-size molecular systems in isolation,[7] the application of these methods to condensed phases is problematic. The poor scaling of the methods generally prohibits their application to large molecules or clusters. Although clever methods have been devised to embed *ab initio* quantum mechanical calculations in classical environments – the so-called QM/MM methods[8] – such methods suffer from certain common difficulties. Firstly, the poor scaling of the quantum 'system of interest' becomes a rate-limiting step which can preclude chemically relevant sampling of configuration space[9]. Secondly, although there are a profusion of methods for treating the boundary between the quantum and classical systems[10], there is as yet no consensus on how to address this boundary even for the ground state – the situation for excited states is arguably worse[11]. Lastly, such methods generally do not have the scope for treating non-classical system-environment interactions. The last difficulty has been surmounted for the case of polarizable continuum environments[12], but at the cost of being able to treat specific interactions such as may be relevant in structured biomolecules or solid-state matrices.

As a result of these difficulties, there has quite often been recourse to simpler techniques based on parameterized model Hamiltonians[13]. These models are often useful for the direct simulation of photochemical phenomena in systems too large or complex to be treated by *ab initio* techniques. Such strategies have been applied to great effect in the description of enzymatic catalysis[14], protein photochemistry[15], the physics of polyenes and aromatics[16], organic reactivity[17], proton transfer problems[18], and the effects of solvent on non-radiative decay[19] and nonlinear optical properties of polyenes[20-22]. In addition, they have been used to great effect in characterizing a common mechanistic behavior in series of otherwise disparate chemical systems[23]. In the latter capacity such models have played an important role in the development of important core chemical concepts[24].

In this paper, we examine a simple model designed to capture features of double bond photoisomerization in molecules like those in figure 1. The defining characteristic of these molecules is a resonance between Lewis structures that differ by bond alteration and the location of a formal charge. They are called *methine dyes*[25]. Such molecules have strong absorption in or near the visible region, which underlies much of their utility[25]. The molecules in figure 1 all display environmentally-dependent fluorescence yields due to competition with a common non-radiative decay mechanism – double bond photoisomerization of the methine bridge[26]. This reaction is associated with decay into conformations with a flipped double bond – decay of Thiazole Orange by this pathway could in principle result in any of the distinguishable products shown in figure 2. The environmental dependence of the competition between fluorescence and photoisomerization makes all of the molecules shown in figure 1 useful as biotechnological markers[27-29]. Similar molecules are also useful as industrial dyes for optical storage media[25] and as model systems for the demonstration of coherent quantum control[30].

Non-radiative decay mechanisms are normally discussed as instances of breakdown of the Born-Oppenheimer approximation (BOA)[31]. Within the BOA, the action of the nuclear kinetic energy on the electrons is disregarded, so that molecular dynamics can be modeled by evolution on potential eigensurfaces of a Hamiltonian that acts on the electronic Hilbert space and is parametrically dependent on the nuclear positions. Such a situation is illustrated in figure 3. The BOA will fail if the energies of the electronic states become too close. In these situations, the nuclear motion can induce transitions between the electronic surfaces.

In general, degeneracies between electronic states have the local topology of a double cone, and are called *conical intersections*[32,33]. These features of the potential energy surfaces are central to mechanistic descriptions of photochemistry, wherein their conceptual importance is comparable to that of transition states in thermochemistry. However, a direct comparison with transition states is misleading – conical intersections are not isolated points in configuration space but multidimensional seams that generally have dimensionality equal to the number of internal degrees of freedom minus two[33]. This means that in the case of ethylene – the simplest molecule that can undergo *Z,E*-photoisomerization – the conical intersection itself is a ten-dimensional manifold. The staggering dimensionality of intersection seams for more complicated systems (fig. 1) is quite difficult to grasp. It represents a long thorn in the side of photochemical pedagogy. It has been relatively common practice to sidestep this by use of *minimal energy conical intersections* (MECIs) as tools of mechanistic deduction[34]. These are single molecular configurations that represent extrema on the intersection seam – usually minimizing the energy on the upper state[35,36]. MECIs can be useful in describing the opening of activated decay processes or excitation threshold phenomena[37], but in the general case their relevance is easily questioned.

Recent studies dedicated towards mapping intersection seams using *ab initio* electronic structure techniques[36] have yielded some interesting results. In several cases it has been demonstrated that previously reported MECIs, which had been associated with distinct decay products, are just different extrema on a single intersection seam[38]. This is more than a curiosity, because the electrons and nuclei become entangled in the vicinity of conical intersections.[39] This means that subsequent observation of the electronic state will produce a statistical mixing of nuclear states, destroying information about their evolution.[40] It is conceivable that if this effect were sufficiently strong, the ability to classically distinguish between different chemical pathways leading to different products would be destroyed. In these cases classical interactions with an environment would not be sufficient to control product yield, and the notion of distinct pathways such as drawn in fig. 3 would become meaningless. Molecules like those in figure 1 are known for the environmental dependence of the yield of fluorescence versus non-radiative decay[29,41,42] – the latter being dominated by double bond photoisomerization processes[43,44]. The existence or nonexistence of classically distinguishable pathways is crucial to understanding the mode of environmental control. In cases where changing the product yield leads to potentially useful photochromism or fluorescence switching, understanding the environmental control can inform the development of useful technologies[45,46].

In this paper we propose a simple parameterized model for the interactions between valence-bond states in a reaction space that describes the photoisomerization of molecules such as in figure 1. The parameterization of the model is simply connected to an approximation to the overlap between fragment frontier orbitals. We examine the behavior of the excited states and conical intersection seams with respect to changes in the parameterization. We will describe general characteristics of the potential surface of the adiabatic states under the interaction and will pay particular attention to the topology and extent of conical intersection seams. The model fulfills the most basic requirements of a double bond photoisomerization model: 1) it describes energetically favorable twisting motion in the excited state and 2) it features conical intersections that can be accessed by this motion and through which non-radiative decay could occur. These intersections, which exist for a range of twist angles, are found along a conrotatory twist pathway. The model predicts intersections between the first and a higher excited state that may be accessed for a larger range of angles – including untwisted configurations. Furthermore, the model suggests the existence of qualitatively different physical regimes wherein intersections between different pairs of states may occur along orthogonal paths or along the same path, and in the latter case describes the induction of barriers on the lower excited state through interaction with the higher one.

**METHODS**

**The Electronic Reaction Space**

Our model is based on a valence-bond ansatz for the electronic structure outlined in Figure 4. We conceive of three orbitals that have $\pi$ character and are localized on three atoms or molecular fragments. The orbitals are labeled $l,b$ and $r$ ("left", "bridge", "right"). Filling the system with either two or four electrons will lead to three "maximally covalent" structures, each featuring a bond and an excess charge. In the four-electron case, these are written as (1).

$$|L_-\rangle = \bar{l}l(b\bar{r} - \bar{b}r)$$
$$|B_-\rangle = b\bar{b}(\bar{l}r - \bar{l}r) \qquad (1)$$
$$|R_-\rangle = r\bar{r}(l\bar{b} - \bar{l}b)$$

In each of these, there is one non-bonding orbital while the remaining orbitals are involved in a bond. In the case of a positive formal charge, two electrons would be distributed over the fragments; appropriate structures can be generated from (1) by deleting the non-bonding electrons. We can also define three *ionic* structures, which, in the four-electron case, are written as (2).

$$|L_+\rangle = b\bar{b}r\bar{r}$$
$$|B_+\rangle = \bar{l}lr\bar{r} \qquad (2)$$
$$|R_+\rangle = \bar{l}lb\bar{b}$$

These are generated by the terms in (1) by moving the electrons in the bond to one of the sites that it joins. In the two-electron case, ionic states can be generated by removing the electrons in one of the doubly filled orbitals from each of the states in (2). It will be convenient to work in a space labeled only by the covalent functions (1). This should be possible if the ionic states (2) lie higher in energy. Then the Hamiltonian can be block-diagonalized to separate covalent and ionic spaces, and it will be meaningful to define an effective three-dimensional Hamiltonian over a space labeled only by the covalent states (1). The states spanning this space will resemble (3).

$$|L\rangle = N_L\left[|L_-\rangle + \sum_I c_I^L|I_+\rangle\right]$$
$$|B\rangle = N_B\left[|B_-\rangle + \sum_I c_I^B|I_+\rangle\right] \qquad (3)$$
$$|R\rangle = N_R\left[|R_-\rangle + \sum_I c_I^R|I_+\rangle\right]$$

Here $I$ indexes the ionic states (2) and the $N$ are normalization constants. The resulting final states are 'perfect-pairing' or 'valence-bond' (VB) states[47,48]. They can be chosen as approximate diabatic states via block diagonalization of the Hamiltonian in the many (in this case, four) body space[49]. We are going to make two assumptions. The first is that a transformation exists which maps the states 3 onto an effective Hilbert space labeled by the covalent functions. This labeling should be meaningful in the sense that projection onto a given covalent state measures the probability of finding the excess charge localized on the fragment to which it is assigned in that state. This will be true if the ionic states contribute only in accordance with their importance to the covalent bonding interaction[50]. Secondly, we will assume a particular parametric form for the interactions between the states in the contracted Hilbert space (below). The parameterization will make the most sense if the transformation preserves the $\pi$ character of the orbitals.

We have tested our assumption of covalent-ionic separation using SA-CASSCF wavefunctions previously described for a green fluorescent protein chromophore anion[43]. We performed a block diagonalization upon the Hamiltonian defined over configuration state functions constructed with Foster-Boys[51] localized orbitals obtained from a three-state-averaged CASSCF calculation[52] with four electrons in three orbitals. The orbitals are identical to the fragment orbitals shown in figure 4. The SA-CASSCF energies and wavefunctions were used to construct a 6 x 6 Hamiltonian that was then block diagonalized using the unitary 'least-action' transformation of Cederbaum et al.[53] The transformed Hamiltonian contained a low-lying 3 x 3 block with leading covalent terms and a high-lying 3 x 3 block with leading ionic terms. The standard deviations of the diagonal energies of the lower and higher block were 0.031 and 0.075 Hartrees (h), and the difference between the mean diagonal energies of the low and high-energy blocks was 0.319 h. The off-diagonal elements of the upper and lower blocks were of the same order of magnitude as the standard deviation of diagonal elements of the blocks. That the splitting between covalent and ionic blocks in the transformed Hamiltonian was an order of magnitude greater than their internal splitting and off-diagonal elements corroborates our assumption of energetic separation between covalent and ionic states for this particular example.

**Parameterized Interaction Model**

Ideas of qualitative valence-bond theory indicate that the interaction matrix elements over our electronic state space can be approximated by orbital overlaps[47,48]. This idea has previously been used in studies of photoisomerization of single π bonds[19,23]. In this paper we will study the properties of the simplified interaction Hamiltonian given in (15).

$$H = \begin{pmatrix} H_{LL} & H_{LB} & H_{LR} \\ \vdots & H_{BB} & H_{BR} \\ \ddots & \dots & H_{RR} \end{pmatrix} = E_0 \begin{pmatrix} 0 & -\cos(\theta_L) & -\xi\cos(\theta_R - \theta_L) \\ \vdots & 0 & -\cos(\theta_R) \\ \ddots & \dots & 0 \end{pmatrix} \quad (4)$$

Here $E_0$ is some appropriate energy scale (on the order of a few eV for molecules in figure 1). The interaction matrix elements of this interaction Hamiltonian are graphically depicted in figure 5. The model is diagonally degenerate and traceless (i.e. we assume that the one-electron and single-site two electron energies of the orbitals are degenerate), and the interaction matrix elements are simple functions of three parameters $\theta_L$, $\theta_R$ and $\xi$. The first two of these are angles and the third is a positive real number. Negative values of $\xi$ do not predict favorable twisting motion on the excited state and therefore do not meet the basic expectations of a photoisomerization model. We will come to this point again later.

**Justification and Interpretation in the Context of a Molecular Frame**

Recalling our conception of the electronic state space (Fig. 4), and referring to the geometrical idealization of such a system drawn in Fig. 5, it is clear that the parameters $\theta_L$ and $\theta_R$ can be readily interpreted as torsion angles. The lower diagonal elements of $H$ can be interpreted as overlaps between the apical sites $l$ and $r$ with the central site $b$. The parameter $\xi$ does not have an immediately clear interpretation in terms of molecular geometry.

In order to better understand what $\xi$ may represent in the context of a molecular frame, we refer to the bottom of figure 5. We imagine our underlying orbital space as three π type orbitals (drawn as hydrogenic p orbitals). Take the π orbital axis vector of the central orbital $b$ to be fixed perpendicular to the molecular plane. Assume also that the lengths of the bonds between the ends and the bridged are symmetric so that $r_{lb} = r_{br}$. The end orbitals $l$ and $r$ to rotate about rigid bonds connecting them to $b$ each (so that the plane swept out by their π orbital axis vectors are perpendicular to the respective bonds). The angles $\theta_L$ and $\theta_R$ measure these rotations in the sense of a left-hand rule for the $\theta_L$ and a right-hand rule for $\theta_R$. We can decompose the overlap between two $p$ orbitals into $\pi$ and $\sigma$ components which are given by formulas (5) and (6) due to Mulliken[54].

$$S_{2p\pi}(\rho_{ab}) = e^{-\rho}[1 + \rho_{ab} + \frac{2}{5}\rho_{ab}{}^2 + \frac{1}{15}\rho_{ab}{}^3] \tag{5}$$

$$S_{2p\sigma}(\rho_{ab}) = e^{-\rho}[-1 - \rho_{ab} - \frac{1}{5}\rho_{ab}{}^2 + \frac{2}{15}\rho_{ab}{}^3 + \frac{1}{15}\rho_{ab}{}^4] \tag{6}$$

These overlaps are different for orbitals belonging to different elements, with the 'atomic type' dependency buried in the quantity (7).

$$\rho_{ab} \equiv \frac{(\mu_a + \mu_b)r_{ab}}{2a_H} \tag{7}$$

Here $r_{ab}$ is the interatomic distance, $a_H$ is the Bohr radius and the $\mu_{a,b}$ are parameters identified with the atom type (element, ionization and hybridization state). The values of $\mu$ for a neutral carbon atom, a carbanion or a carbocation are 1.625, 1.45 and 1.80, respectively. For the four-electron, three-site case (fig. 4), we may use a 2:1 carbon carbanion average for the $\mu$'s ($\mu$=1.57) which corresponds to $\rho_{lb} = \rho_{br}$ = 2.97Å$^{-1}r_{lb}$. There is only one bond length of interest for this simple example because $r_{lr} = 2r_0 sin(\phi/2)$ with $\phi$ being the interbond angle – we drop the subscripts and define $\rho = \rho_{lb} = \rho_{br}$. Now $r \sim 1.45$ Å for a resonant carbon-carbon bond so a reasonable value for an allylic system would be $\rho = 4.37$.

The total overlap will be expressed as a linear combination of the terms (5) and (6) weighted by products of direction cosines of the $\pi$ orbital axis vectors. The overlaps between the sites in the geometrical model are then given by (8).

$$S_{lb} = S_{2p\pi}(\rho)\cos(\theta_L) \tag{8a}$$

$$S_{br} = S_{2p\pi}(\rho)\cos(\theta_R) \tag{8b}$$

$$\begin{aligned}
S_{lr} &= S_{2p\pi}(2\rho\sin(\phi/2))\cos(\theta_L)\cos(\theta_R) \\
&\quad + [S_{2p\sigma}(2\rho\sin(\phi/2))\sin^2(\phi/2) + S_{2p\pi}(2\rho\sin(\phi/2))\cos^2(\phi/2)]\sin(\theta_L)\sin(\theta_R) \\
&= \frac{1}{2}\Big[S_{2p\pi}(2\rho\sin(\phi/2)) + S_{2p\sigma}(2\rho\sin(\phi/2))\sin^2(\phi/2) + S_{2p\pi}(2\rho\sin(\phi/2))\cos^2(\phi/2)\Big]\cos(\theta_R - \theta_L) \\
&\quad + \frac{1}{2}\Big[S_{2p\pi}(2\rho\sin(\phi/2)) - S_{2p\sigma}(2\rho\sin(\phi/2))\sin^2(\phi/2) - S_{2p\pi}(2\rho\sin(\phi/2))\cos^2(\phi/2)\Big]\cos(\theta_R + \theta_L)
\end{aligned} \tag{8c}$$

In the last part of (8c) we have used the trigonometric identities $cos\alpha\,cos\beta = [cos(\alpha-\beta) + cos(\alpha+\beta)]/2$ and $sin\alpha\,sin\beta = [cos(\alpha-\beta) + cos(\alpha+\beta)]/2$.

We can then define quantities $\xi$ and $\delta$ as in (9).

$$\delta \equiv \frac{\left[ S_{2p\pi}\big(2\rho\sin(\phi/2)\big) - S_{2p\sigma}\big(2\rho\sin(\phi/2)\big)\sin^2(\phi/2) - S_{2p\pi}\big(2\rho\sin(\phi/2)\big)\cos^2(\phi/2) \right]}{2S_{2p\pi}(\rho)} \qquad (9)$$

We can now write the $l$-$r$ Hamiltonian as (10).

$$H = S_{2p\pi}^{-1}(\rho)\begin{pmatrix} 0 & -S_{lb} & -(S_{lr} - \delta S_{2p\pi}(\rho)\cos(\theta_R + \theta_L)) \\ \vdots & 0 & -S_{br} \\ \ddots & \dots & 0 \end{pmatrix} \qquad (10)$$

Comparing this with (15) again, we see that, in this example, the parameter $\xi$ takes the form (11) in the present example.

$$\xi \equiv \frac{\left[ S_{2p\pi}\big(2\rho\sin(\phi/2)\big) + S_{2p\sigma}\big(2\rho\sin(\phi/2)\big)\sin^2(\phi/2) + S_{2p\pi}\big(2\rho\sin(\phi/2)\big)\cos^2(\phi/2) \right]}{2S_{2p\pi}(\rho)} \qquad (11)$$

Thus, $\xi$ may be understood as a particular component of the ratio of the overlap of sites $l$ and $r$ relative to the overlap of each of these with $b$. It is a nonlinear two-to-one function of the parameters $\rho$ and $\phi$ in our geometrical idealization. It can readily be visualized by contours in the plane spanned by $\rho$ and $\phi$, as we present in figure 6.

In this paper, we are going to examine the simplified situation where terms corresponding to $\delta$ are neglected. The terms corresponding to $\delta$ should be generally less than that of $\xi$ because the fragment orbitals can be chosen so that the overlaps are positive at bonding distances. This implies that $\delta$ represents a difference of positive numbers while $\xi$ represents a sum of the same numbers. We are guaranteed then that $|\xi| \geq |\delta|$. The dependence of the terms on $\cos(\theta_R$-$\theta_L)$ and $\cos(\theta_R$+$\theta_L)$ means that this approximation will fail when $\cos(\theta_R$-$\theta_L) = 0$ and $\cos(\theta_R$+$\theta_L) \neq 0$. We have investigated the situation using the value for $\rho$ suggested above and $\phi = 2\pi/3$ as appropriate for $sp^2$ carbon and found that this is true only in limited regions of the plane spanned by $\theta_L$ and $\theta_R$. We will examine what happens when $\delta$ is added as an independent parameter in an upcoming paper, at which point we will benefit from having thoroughly characterized the limit where $\delta = 0$. Of course, results for the limit of $\delta > 0$ and $\xi = 0$ can be extracted from the present results by the map $\theta_L \rightarrow -\theta_L$.

We anticipate that the approximation of neglect of $\delta$ will become more robust when one considers molecules such as in fig. 1 rather than allylic systems (which fig. 5 most clearly invokes). The reason is that when the end groups are not atoms but planar aromatic rings, there is also substantial overlap between the atoms *ortho* to the bridge connection at planar configurations that is removed when the bonds are twisted. We therefore argue that, for a planar aromatic ring, the overlap components multiplying $\cos(\theta_R)\cos(\theta_L)$ in equation (8c) should be larger than the components which multiply $\sin(\theta_R)\sin(\theta_L)$. This argument is supported by the observation that cyclization dominates the photochemistry of unsubstituted allyl anions[55] while double bond isomerization is observed aryl-substituted allyl anions[56]. Although the right hand side of equation (8c) could have been further reduced algebraically, we chose not to do in order to highlight the potential for future generalizations where the $S_{2p\pi}$ overlaps multiplying $\cos(\theta_R)\cos(\theta_L)$ and $\sin(\theta_R)\sin(\theta_L)$ are used as independent parameters that characterize the shape of a molecular fragment.

The exposition above makes clear that, in order to achieve a reduced dimensionality in our parameterization, we have placed certain symmetry constraints on its representation of an actual molecule. It is important to enumerate them because in future works we will study what happens when they are removed. These symmetries are of two types. The first is the symmetry invoked by stating that the diagonal energies of the VB states are the same. In general, fragments on a methine-bridged molecule will have different one-electron (orbital) energies as well as different single-fragment and multi-fragment correlation contributions – although we are interested primarily in cases where the spread of diagonal energies is less than the couplings (the resonant case). The second is the symmetry inherent in the couplings, which can be interpreted geometrically as an equivalence of the bonds between the ends and the bridge. Note that there are no parameters whose action affects multiple sites asymmetrically through either the diagonal or off diagonal elements (the angle difference in the higher off-diagonal element is symmetric in its action because of the parity of the cosine function). The geometric interpretation of this limit is that we are constraining the methine bonds to be equal. We will investigate the effects associated with freeing these degrees of freedom in future work, at which point we will benefit from having thoroughly characterized the present limit.

**Expressions for the Eigenenergies and Eigenstates**

The condensation of the electronic structure space to three states from six allows the use of analytic solutions for the eigenvectors $(\nu_0, \nu_1, \nu_2)$ and eigenenergies $(\varepsilon_0, \varepsilon_1, \varepsilon_2)$ of a 3x3 symmetric matrix $H$. Following Cocolicco and Viggiano[57], we define the quantities (4) – (6).

$$b = -Tr(H) \tag{12}$$

$$c = \frac{1}{2}\left[\left(TrH\right)^2 - Tr(H^2)\right] \tag{13}$$

$$d = -Det(H) \tag{14}$$

These represent first through third order traces over the matrix. We then define parameters (15) – (17).

$$u = -\frac{1}{9}\left(b^2 - 3c\right) \tag{15}$$

$$v = \frac{1}{54}\left(2b^3 - 9bc + 27d\right) \tag{16}$$

$$\cos\Theta = -\left(\frac{v}{u\sqrt{|u|}}\right) \tag{17}$$

For cases such as (4), where the average of the diagonal elements is zero, the eigenvalues are then given by (19).

$$\varepsilon_0 = -2\sqrt{u}\cos\left(\frac{\Theta}{3}\right) \tag{19a}$$

$$\varepsilon_1 = \sqrt{|u|}\left[\cos\left(\frac{\Theta}{3}\right) - \sqrt{3}\sin\left(\frac{\Theta}{3}\right)\right] \tag{19b}$$

$$\varepsilon_2 = \sqrt{|u|}\left[\cos\left(\frac{\Theta}{3}\right) + \sqrt{3}\sin\left(\frac{\Theta}{3}\right)\right] \tag{19c}$$

The eigenvectors are given by (20)

$$\vec{v}_{i \in \{0,1,2\}} = \begin{pmatrix} -(\varepsilon_i - H_{22})H_{13} - H_{12}H_{23} \\ -(\varepsilon_i - H_{11})H_{23} - H_{21}H_{13} \\ -(\varepsilon_i - H_{11})(\varepsilon_i - H_{22}) + H_{12}^2 \end{pmatrix} \qquad (20)$$

Where $H_{ij}$ is the element in the $i$th row and $j$th column of $H$. In more general cases the $\varepsilon$ of (19) and (20) are eigenvalues shifted by the diagonal average. Note that although the symmetri[58]c matrix possesses five independent elements, the energy splittings are completely determined by only two functions of these: $u$ and $v$ (or equivalently $u$ and $\Theta$). We will always number the states in increasing order by energy. We will refer to the states by the usual notation for electronic singlets: $S_0$, $S_1$, $S_2$. The formulas above are valid for any symmetric positive-definite 3 x 3 matrix and do not depend on spin or any other physically relevant quantity used in the parameterization.

The above formulas are valid on a manifold-with-boundary defined by $u \leq 0$ and $v^2 + u^3 \leq 0$. On this manifold-with-boundary eigenvalues of the matrix must be real. On the manifold itself, where the inequalities are strong, the eigenvalues are non-degenerate. On the boundary of the manifold at least two of the eigenvalues become degenerate. On the sub-manifold $v > 0$ of the boundary, $\Theta = \pi$ and $\varepsilon_0 = \varepsilon_1$. On the sub-manifold $v < 0$, $\Theta = 0$ and $\varepsilon_1 = \varepsilon_2$. At the point $(u,v)=(0,0)$ the problem is completely singular, $\Theta$ is not a number and three-fold degeneracy arises. These conditions can be summarized by the condition (21).

$$v = \pm u\sqrt{|u|} \qquad (21)$$

Here the relation is positive for $S_1/S_0$ degeneracy, negative for $S_2/S_1$ degeneracy and both sides vanish for a 3-state degeneracy. If one differentiates both sides of the equation with respect to some underlying parameterization (as above), one obtains (22).

$$\nabla v = \pm \frac{3\sqrt{|u|}\nabla u}{2} \qquad (22)$$

This equation must be satisfied component-wise in the tangent space on the boundary. The boundary represents the union of all degeneracies that are possible within the model, regardless of any mapping onto an underlying parameterization. The different signs arise in a manner identical to (21). A graphical illustration of the eigenvalues given as functions over the (u,v) plane is displayed in figure 7, as well as an illustration of the domain that gives real eigenvalues. The boundary of the manifold is highlighted. Note that it is continuous, but that its derivatives in the (u,v) plane will not be.

If we apply our parameterized model to the expressions for the quantities u and v (eqns. 15 & 16) gives

$$u = -\frac{1}{3}\left(\cos^2(\theta_L) + \cos^2(\theta_R) + \xi^2 \cos(\theta_R - \theta_L)\right) \tag{23}$$

and

$$v = \xi \cos(\theta_L)\cos(\theta_R)\cos(\theta_R - \theta_L) \tag{24}$$

These quantities are then sufficient to specify all the energy splittings of our model defined over the three-dimensional ($\theta_L$, $\theta_R$, $\xi$) parameter space. Note that they are continuous, but not single-valued functions of $\theta_L, \theta_R$ and $\xi$. The topology of the intersection space will reflect the multi-valued nature of $u$ and $v$.

Substitution of the matrix elements in (4) into the eigenvector expression (20) gives (25).

$$\vec{v}_{i \in \{0,1,2\}} = \begin{pmatrix} \varepsilon_i \xi \cos(\theta_R - \theta_L) - \cos(\theta_L)\cos(\theta_R) \\ \varepsilon_i \cos(\theta_R) - \xi \cos(\theta_L)\cos(\theta_R - \theta_L) \\ -\varepsilon_i^2 - \cos^2(\theta_L) \end{pmatrix} \tag{25}$$

**The ($\theta_L$, $\theta_R$) Plane**

We will discuss in some detail the behavior of the eigenvalues (19) over the plane spanned by the angles $\theta_L$ and $\theta_R$. In this discussion, it will be convenient to refer to particular coordinates in that plane. These coordinates are indicated in Fig 8, which depicts a "unit cell" of the plane ($0 \le \theta_1 < \pi, 0 \le \theta_2 < \pi$). Due to the periodicity of the functions used in our parameterization, all other points in the ($\theta_L, \theta_R$) plane are periodic images of points in this cell. The figure also highlights important directions in the plane to which we will refer. We will make reference to coordinates that correspond to twisting one bond while the other remains constant as "one-bond twists". Note that these coordinates have an easy interpretation as coordinates of a single π bond. We will also make reference to even and odd combinations of single bond twisting – conrotatory twisting and disrotatory twisting of the bonds. These coordinates are schematically illustrated in figure 8. These are quantitatively defined as in (26).

$$\theta_C = \theta_L - \theta_R$$
$$\theta_D = \theta_R + \theta_L$$
(26)

The parity of the combinations in (26) is dictated by our choice of left-hand and right-hand rules to define the coordinates $\theta_L$ and $\theta_R$, respectively. If we had used a single-handed sense to define both rotations, then the parities would be reversed. This will not affect the physics however, and it is *much easier* to keep track of which coordinate is which by thinking of the symmetry of a model like that in figure 5. If the molecular frame has $C_{2v}$ symmetry when no twist is applied, then *the conrotatory twist preserves $C_2$ symmetry while breaking $C_s$ and the disrotatory twist preserves $C_s$ symmetry while breaking $C_2$*. Here, the plane defining $C_s$ symmetry is that which perpendicularly bisects the molecule through the bridge.

The conrotatory and disrotatory twist coordinates are *emergent* coordinates because they have no meaning in a discussion of a single π bond, emerging instead from the joint coordinate space of two π bonds coupled together. When we refer to a "planar configuration" of our model, we mean one of the edges of the corners of the unit cell in Fig 8, i.e. (0,0), (0,π), (π,0) or (π,π).

The photoisomerization of alkene and methine molecules is sometimes discussed in the context of the so-called "hula-twist"[59]. This mechanism has been, for example, invoked in the context of fluorescent protein chromophore photoisomerization[60]. It is usually invoked in situations where the end groups themselves are constrained[59], such as may occur for a chromophore bound within a protein. The central point is that photoisomerization may be accomplished by motion of the bridge alone. The path taken by the bridge in the rest frame of the stationary rings is taken to be analogous to the path followed by a hula-dancer's hips in the rest frame of his (or her) head and feet. In the context of the geometric interpretation of our model given above, it is a form of disrotatory twisting motion. The difference between the usual algebraic formula for the conrotatory twist and ours for disrotatory twist (eqn. 26) is again a consequence of the handedness we have chosen to define $\theta_L$ and $\theta_R$.

**Derivative Expressions**

As mentioned above, the quantities $u$ and $v$ determine the energies of all the states of our model to within a constant shift, and therefore completely determine the energy splittings, gradients and degeneracy spaces. This led to the equations (22), which must be satisfied componentwise within the intersection spaces. When we consider the topology and range of the intersection spaces, it will be useful to have at hand formulas for the derivatives of these quantities with respect to the parameter space. Differentiating with respect to the original parameter set $(\theta_L, \theta_R, \xi)$, we obtain the expressions (27) and (28).

$$
\begin{aligned}
\frac{\partial v}{\partial \theta_L} &= -\xi\big(\sin\theta_L \cos\theta_R \cos(\theta_R - \theta_L) - \cos\theta_L \cos\theta_R \sin(\theta_R - \theta_L)\big) \\
&= -\xi \cos\theta_R \sin(\theta_R + 2\theta_L) \\
\frac{\partial v}{\partial \theta_R} &= -\xi\big(\cos\theta_L \sin\theta_R \cos(\theta_R - \theta_L) + \cos\theta_L \cos\theta_R \sin(\theta_R - \theta_L)\big) \\
&= -\xi \cos\theta_L \sin(2\theta_R - \theta_L) \\
\frac{\partial v}{\partial \xi} &= \cos\theta_L \cos\theta_R \cos(\theta_R - \theta_L)
\end{aligned}
\tag{27}
$$

$$
\begin{aligned}
\frac{\partial u}{\partial \theta_L} &= \frac{1}{3}\big(2\sin\theta_L \cos\theta_L - 2\xi^2 \sin(\theta_R - \theta_L)\cos(\theta_R - \theta_L)\big) \\
&= \frac{1}{3}\big(\sin(2\theta_L) - \xi^2 \sin(2\theta_R - 2\theta_L)\big) \\
\frac{\partial u}{\partial \theta_R} &= \frac{1}{3}\big(2\sin\theta_R \cos\theta_R + 2\xi^2 \sin(\theta_R - \theta_L)\cos(\theta_R - \theta_L)\big) \\
&= \frac{1}{3}\big(\sin(2\theta_R) + \xi^2 \sin(2\theta_R - 2\theta_L)\big) \\
\frac{\partial u}{\partial \xi} &= -\frac{1}{3}\big(\cos^2\theta_L + \cos^2\theta_R + 2\xi \cos(\theta_R - \theta_L)\big)
\end{aligned}
\tag{28}
$$

It will actually prove more convenient to have transformed these quantities into the equivalent parameter set $(\theta_C, \theta_D, \xi)$. With respect to this set, $u$ and $v$ are expressed as (29).

$$v = \xi \cos\left(\frac{\theta_D + \theta_C}{2}\right) \cos\left(\frac{\theta_D - \theta_C}{2}\right) \cos\theta_C$$

$$= \xi \cos\theta_C \left(\cos^2\left(\frac{\theta_D}{2}\right)\cos^2\left(\frac{\theta_C}{2}\right) - \sin^2\left(\frac{\theta_D}{2}\right)\sin^2\left(\frac{\theta_C}{2}\right)\right) \qquad (29)$$

$$u = -\frac{1}{3}(\cos^2\frac{\theta_D + \theta_C}{2} + \cos^2\frac{\theta_D - \theta_C}{2} + \xi^2 \cos^2\theta_C)$$

The appropriate derivative expressions are (30) and (31).

$$\frac{\partial v}{\partial \theta_D} = -\frac{\xi}{2}\cos\theta_C \sin\theta_D$$

$$\frac{\partial v}{\partial \theta_C} = -\frac{\xi}{2}\cos\theta_D \sin\theta_C - \xi\cos\theta_C \sin\theta_C$$

$$\frac{\partial v}{\partial \xi} = \cos\left(\frac{\theta_D + \theta_C}{2}\right)\cos\left(\frac{\theta_D - \theta_C}{2}\right)\cos\theta_C \qquad (30)$$

$$= \cos\theta_C\left(\cos^2\left(\frac{\theta_D}{2}\right)\cos^2\left(\frac{\theta_C}{2}\right) - \sin^2\left(\frac{\theta_D}{2}\right)\sin^2\left(\frac{\theta_C}{2}\right)\right)$$

$$\frac{\partial u}{\partial \theta_D} = \frac{1}{6}\left(\sin(\theta_D + \theta_C) + \sin(\theta_D - \theta_C)\right)$$

$$\frac{\partial u}{\partial \theta_C} = \frac{1}{6}\left(\sin(\theta_D + \theta_C) - \sin(\theta_D - \theta_C) + 2\xi^2 \sin(2\theta_C)\right) \qquad (31)$$

$$\frac{\partial u}{\partial \xi} = -\frac{1}{3}(\cos^2\frac{\theta_D + \theta_C}{2} + \cos^2\frac{\theta_D - \theta_C}{2} + 2\xi\cos^2\theta_C)$$

## RESULTS

### Potential surfaces over ($\theta_L$, $\theta_R$) plane

#### The Critical Level $\xi = 0$

Figure 9 displays global views of the potential energy surfaces over the ($\theta_L$,$\theta_R$) plane in the case where $\xi = 0$. In this case, there is no interaction between the sites $l$ and $r$. At the point where both $\theta_L$ and $\theta_R$ are equal to $\pi/2$, all the elements of $H$ are zero and the model is singular and degenerate. A three state intersection arises at this point. The $S_0$ and $S_2$ states are evenly split about $S_1$ at all points in the plane, and the $S_1$ energy is not dependent on position in the plane.

#### The Regime $0 < \xi < 1$

For $\xi \neq 0$ there is no 3-state intersection in the ($\theta_L$,$\theta_R$) plane. Instead, separate two-state $S_0/S_1$ and $S_1/S_2$ intersections arise. For each pair of states in the regime $0 < \xi < 1$, two intersections arise in the plane and move outward from the centre along different coordinates. The $S_1/S_0$ intersections separate along the conrotatory twist direction and the $S_2/S_1$ intersections separate along the disrotatory twist direction. A representative situation in the regime where $0 < \xi < 1$ (specifically, $\xi = 0.5$) is included in Fig. 10.

The topology of the $S_1$ surface has developed more structure relative to the $\xi = 0$ case. Planar configurations represent degenerate saddle points on the $S_1$ surface, so that bond twisting is generally favorable from planar configurations. The gradient vanishes at planar configurations; the most likely path taken by the system if excited to the $S_1$ state would be along the directions of greatest negative curvature, along which the magnitude of the gradient (classical force) accumulates most rapidly. Within this regime, the direction of greatest negative curvature on the $S_1$ state at the planar configurations is the conrotatory twist. The curvature in the disrotatory twist direction is zero, so that the $S_1$ surface is flat in this direction up until the $S_2/S_1$ intersection is reached, whereupon further motion in this direction lowers the $S_1$ energy. There is a saddle point on the $S_1$ surface in the center of the unit cell of the plane, with positive curvature along the disrotatory twist and negative curvature along the conrotatory twist.

The $S_2$ surface has a local maximum at planar configurations, so that bond twisting is favorable in all directions. The direction of greatest curvature is the disrotatory twist, which leads to the $S_2/S_1$ intersection. Beyond the intersection, the $S_2$ surface is flat in this direction.

The $S_0$ surface has local minima at the planar configurations, so bond twisting is disfavored on this surface. The direction of greatest positive curvature is the conrotatory twist. The directions of least positive curvature are the one-bond twists. The curvature along the disrotatory twist is intermediate between the two.

### *The Critical Level $\xi = 1$*

The situation at $\xi = 1.0$ represents a critical level in the model which separates qualitatively different regimes. The situation is illustrated in figure 11. At this level, the pair of $S_2/S_1$ which moved outwards towards the corners of the unit cell along the disrotatory twist coordinate have reached (0,0) and $(\pi, \pi)$. Inspection of the potential surface and gradients at these points indicates that the intersections no longer have a conical topology – they have become glancing, or Renner-Teller, intersections[32]. The $S_1/S_0$ intersections remain near the center of the plane and are conical, but continue to move outward along the conrotatory twist coordinate at a slower rate of change with respect to $\xi$ than the $S_1/S_2$ coordinate, which has already moved twice the distance over the same interval of $\xi$. The periodicity of the $(\theta_L, \theta_R)$ plane implies that at this point the $S_2/S_1$ intersection seams, having reached the corners of the unit cell, have "collided" with the incoming seams from the diagonally opposed unit cell.

At this level, the $S_2$ surface is completely flat and featureless. There are local maxima at planar configurations on the $S_1$ surface, so that bond twisting is favorable in all directions. There is a saddle point on the $S_1$ surface in the center of the unit cell, which has negative curvature in the conrotatory twist direction and positive curvature in the disrotatory twist direction. The $S_0$ surface has local minima at planar configurations, so that twisting is generally disfavored. The direction of greatest positive curvature is the conrotatory twist, and the direction of least positive curvature is the disrotatory twist.

### The Regime $1 < \xi < \infty$

For increasing values of $1 < \xi < \infty$, a new regime emerges. The situation is illustrated in figure 12. Within this regime, the $S_2/S_1$ and $S_1/S_0$ degeneracies occur along the *same* coordinate in the plane – the conrotatory twist coordinate. This implies that the intrinsic possible dimensionality of the multi-state decay dynamics has been reduced, and motion along a single coordinate now allows access to both $S_2/S_1$ and $S_1/S_0$ intersections. At the same time, the topography of the $S_1$ state has changed and the curvature in the conrotatory twist direction is now positive, leading to activated ascent towards the $S_2/S_1$ intersection. The curvature of the $S_1$ surface is now negative along the disrotatory twist coordinate. In this regime there are still saddle points at $(\pi/2, \pi/2)$ on the $S_0$ and $S_1$ surfaces, and the curvature along the disrotatory and conrotatory twists maintain the same signs as in earlier regimes. The $S_2$ surface displays a saddle point with negative curvature along the conrotatory twist coordinate and zero curvature in the disrotatory twist coordinate.

As $\xi$ continues to be increased, the position of the $S_2/S_1$ intersection now moves inward along the conrotatory twist coordinate toward the center of the unit cell while the $S_1/S_0$ intersection moves outward along the same coordinate. As $\xi$ approaches infinity the intersections will asymptotically converge at limiting values of $(\pi/4, 3\pi/4)$ and $(3\pi/4, \pi/4)$ in the plane.

**Potential surfaces over a plane spanned by $\xi$ and an angle**

Figure 13 contains illustrations of surfaces formed by the eigenvalues of the interaction Hamiltonian (15) over a plane spanned by $\xi$ and various directions in the $(\theta_L, \theta_R)$ plane. The regimes mentioned above can be seen in the figure. In particular, in the regime $0 < \xi < 1$, the gradient of the $S_1$ energy with respect to $\xi$ is negative while the $S_2$ gradient is positive. The situation reversed for the regime $\xi > 1$. Recall that the geometric interpretation of $\xi$ implies that decreasing $\xi$ corresponds to a combination of internal angle and symmetric bond contraction, so that the model predicts bond and angle expansion upon excitation to $S_1$ at planar configurations in the regime $0 < \xi < 1$. In most cases, the degeneracy at the intersection seams is preserved along curvilinear combinations of $\xi$ and a direction in the $(\theta_L, \theta_R)$ plane.

**Topography and topology of the intersection seam**

Conical intersections between adiabatic potential energy surfaces occur in *seams* (hyperlines) through (generally highly multidimensional) parameter spaces[33]. One of the benefits of working with a reduced three-dimensional model is that the seam can be visualized in its entirety. Regardless of the dimensionality of the underlying parameter space, the locations of conical intersection seams in a three-dimensional Hilbert space are *entirely determined by u and v* (eqns 15 and 16). It follows that the high dimensionality of conical intersection seams in problems involving three coupled electronic states arises solely from the many-to-one mapping of onto the parameter space onto $u$ and $v$. In our model, the parameter space itself is only three-dimensional, and so can be visualized by a human being. Figure 14 displays surfaces of constant $|u|$-$|v|$, which form tubular surfaces enclosing the length of the conical intersection seam. The entire $(\theta_L, \theta_R)$ plane is tiled with periodic images of the conical intersection seam, in accord with the periodic mapping of $u$ and $v$ onto the parameters $\theta_L$ and $\theta_R$.

The behavior of the quantities $u$ and $v$ (eqns 15 & 16) is key to understanding the behavior of the intersection seam. Conical intersections between pairs of states occur when the angle $\Theta$ (equation 9) assumes values of 0 ($S_2/S_1$ intersection) or $\pi$ ($S_1/S_0$ intersection). $H$ has real eigenvalues, so it is always true that $u \leq 0$ along the seam, with equality occurring only at a three-state intersection. Accordingly, regions of the parameter space can be partitioned according to the sign of $v$, and this partitioning will determine conclusively where intersections between distinct pairs of states can occur. Figure 14 displays a tiling of the unit cell of the $(\theta_L, \theta_R)$ plane, with regions of positive $v$ shaded. As can be seen by comparison with the contour surfaces around the seam on the unit cell, the tiling does indeed dictate the regions in which different pairs of states intersect. The partition shown is for $\xi > 0$. The sign of $v$ in the distinct regions will reverse for $\xi < 0$ because $v$ is linear in $\xi$.

It is clear at this point that our model accommodates *one* intersection seam, in the sense that all points of conical intersection are topologically connected modulo the periodicity of the functions used in the parameterization. This is actually clear if one considers Figure 7, and subsequently recognizes that our interaction model contains only continuous functions of the parameters. The quantities $u$ and $v$ are continuous in the matrix elements, being polynomials of these. If the intersection space over the $(u,v)$ seam is continuous, then it must also be so over the parameter space[61]. The branches corresponding to different pairs of intersecting states are distinct for $\xi \neq 0$ but connect at $\xi = 0$, forming a three-state intersection. Surfaces of constant $\xi$ in the full parameter space intersect one pair of $S_1/S_0$ intersections for all $\xi \neq 0$ per unit cell and one three-state intersection per unit cell at the critical level $\xi = 0$.

Although the intersection seam is continuous, its tangent space is not. From consideration of Figure 5 alone, one can expect a discontinuity in the tangents to the seam where the 3-state intersection occurs at $(\theta_L, \theta_R, \xi) = (\pi/2, \pi/2, 0)$. This point is a critical level with respect to $u$ and $v$ on the seam; the value of $\Theta$ becomes undefined at this point because both $u$ and $v$ are simultaneously zero. Here the $S_2/S_1$ and $S_1/S_0$ intersections coalesce.

The origin of the tangent discontinuity at $\xi = 1$ is less clear, and cannot be inferred visually from the plot of the eigenvalues over $(u,v)$ in figure 7. At this level of $\xi$ the $S_2/S_1$ branches approaching from diagonally adjacent unit cells along the disrotatory twist coordinate 'collide' and 'scatter' into the conrotatory coordinate. As can be seen in figure 10, they cease to be conical intersections at this point and become glancing (Renner-Teller) intersections[32]. This phenomenon has been observed before in *ab initio* calculations and has been referred to as 'pair annihilation' of conical intersections[62].

The change in direction of the $S_2/S_1$ branch of the seam can be deduced from the conditions on the derivatives of $u$ and $v$ in the intersection space (eqn. 14). In this case the sign is negative, and equation (22) can be rearranged to give equation (32).

$$\nabla v + \sqrt{|u|}\nabla u = 0 \tag{32}$$

Equation (27) must be satisfied component-by-component in the $S_2/S_1$ intersection space. Taking the derivatives along the lines $\theta_C = 0$ and $\theta_D = \pi$, which are the paths taken by the $S_2/S_1$ branch in the $0 < \xi < 1$ and $\xi > 1$ regimes, and using some algebraic reduction, we obtain the conditions (33).

$$\left.\frac{\partial v}{\partial \theta_D}\right|_{\theta_C = 0} + \frac{3}{2}\sqrt{|u|}\left.\frac{\partial u}{\partial \theta_D}\right|_{\theta_C = 0} = -\xi + \sqrt{\left|-\frac{1}{3}(\cos\theta_D + \xi^2 + 1)\right|} = 0 \tag{33a}$$

$$\left.\frac{\partial v}{\partial \theta_C}\right|_{\theta_D = \pi} + \frac{3}{2}\sqrt{|u|}\left.\frac{\partial u}{\partial \theta_C}\right|_{\theta_D = \pi} = \xi - \frac{\xi^2}{2}\sqrt{\left|-\frac{1}{3}(\cos\theta_C + \xi^2\cos^2\theta_C + 1)\right|} = 0 \tag{33b}$$

Equation 33a can only be satisfied for $\xi \le 1$. Equation 33b can only be satisfied if $\xi \ge 1$. This means that for $\xi \le$ the tangent on the seam can only have a component along $\theta_D$, and for $\xi$ it can only have a component along $\theta_C$.

**Character of the Adiabatic States**

***Charge Localization At Twisted Configurations***

One of the common trends observed in molecules such as those in Fig. 1 is a correlation between twisting and charge separation in their excited state[43,63-67]. In fact, charge separation seems to be a general feature of double bond photoisomerization, even for species with uncharged ground states.[34]

Figure 15 displays the probability of observing the separate charge-localized basis states (eq. 3) over the adiabatic surfaces of our model, distributed over the plane $(\theta_L, \theta_R)$ plane. These are obtained by squaring the components of the adiabatic state vectors onto the VB states of the basis.

The distribution of population of the VB states varies considerably depending on location in the plane, and the extent of mixing differs for the different adiabatic states. The localization is generally highest on the $S_1$ state, as reflected in the saturation of the colors in the plane. Conversely, the muted colors of the $S_0$ and $S_2$ plots reflect greater mixing. The parity of the superpositions represented in the state vectors will be different for these states, as they are orthogonal.

The statistical distribution of the different VB states can be categorized according to the same division of regimes that we have observed in other aspects of the model. Specifically the regimes separated by the critical levels of $\xi$ at $\xi = 0$ and $\xi = \pm 1$. We now outline the characteristics of the distributions in the different regimes and at the critical levels separating them.

### *The Critical Level at $\xi = 0$*

At the critical level at $\xi = 0$, there is no population of the VB state $|B>$ in the $S_1$ adiabatic state at any point in the $(\theta_L, \theta_R)$ plane. Population of this VB state is found only on the $S_0$ and $S_2$ surfaces. The $S_0$ and $S_2$ surfaces display indistinguishable populations at this critical level – the parities of the superposition of VB states in these eigenstates are in opposition as required by orthogonality. Generally, the parity of $S_0$ goes as $|L> + |B> + |R>$ at this level of $\xi$ while the parity of $S_1$ goes as $|L> - |R>$ and the parity of $S_2$ is as $|L> - |B> + |R>$. In the language of valence-bond excited states, they are "twin states" arising from the inversion of a resonance between the VB state $|B>$ and a contracted VB state proportional to $(|L> + |R>)$.[68,69] The $S_1$ state, on the other hand, is in some sense a twin state of *both* $S_0$ and $S_2$ that arises from the inversion of the parity of the contracted state $(|L> + |R>)$ to produce the anti-resonant state (|L> - |R>). The relative weight of each goes approximately as the degree of progress along the different one-bond twists taken independently. If one of the angles is increased keeping the other at zero, the effect is to concentrate one of the end-localized VB states in the $S_1$ adiabatic state to the exclusion of the other while the remaining state is mixed more heavily into the $S_0$ and $S_2$ states. The state that is removed from S1 is the state with single-bond character on the twisted bond. When both bonds are planar, the |L> and |R> VB states are equally populated in all states.

### *The Regime $0 < \xi < 1$*

As $\xi$ is increased in the regime $0 < \xi < 1$, more structure develops. While in the region near the edges of the $(\theta_l,\theta_r)$ unit cell the situation is similar to that at $\xi=0$, the distribution of charge states changes dramatically in a rhomboid region in the center of the unit cell defined by the bifurcating $S_2/S_1$ and $S_1/S_0$ conical intersections which occur in this regime.

Outside the rhomboid region the centre-localized VB state |B> projects mostly onto $S_0$ and $S_2$. The $S_0$ wavefunctions go as $a$|L> + $b$|B> + $c$|R> in this region, with $b > a,c$. The form of the $S_2$ state vector is similar, but with the parity of |B> reversed with respect to |L> and |R>. The $S_1$ state vector goes as |L> - |R> to a good approximation. Within the rhomboid region, the situation abruptly changes – the population of |B> becomes concentrated in the $S_1$ state While $S_0$ and $S_2$ become superpositions of |L> and |R>. The change is most abrupt in the vicinity of the conical intersections, where the populations change discontinuously. In a neighborhood of any of the four conical intersections on the $S_1$ state, it is possible to sample regions dominated almost exclusively by each of the three charge-localized structures. For the $S_0$ and $S_2$ states a similar situation exists for the two intersections involving the state in question, while in the region of intersections not involving that state the charge distribution is extensively mixed.

### *The Critical Level at $\xi = 1$*

At the critical level of $\xi = 1.0$, the $S_2/S_1$ intersection pair has reached the (0,0) and $(\pi,\pi)$ corners of the $(\theta_L,\theta_R)$ unit cell. In the vicinity of these configurations, small changes in position in the plane bring about large changes in the population of the VB states. If one imagines Frank-Condon excitation of an equilibrated state at (0,0), the resulting excited state wave packet would sample zones dominated almost exclusively by each of the VB states. A similar situation holds for the $S_2$ state near the planar $S_2/S_1$ intersection, while the ground state at planar configurations is extensively mixed in a neighborhood of any of the corners of the unit cell. Large changes to this state and $S_1$ persist in a neighborhood of the $S_0/S_1$ intersection on the interior of the unit cell along the conrotatory twist coordinate.

### *The Regime $1 < \xi < \infty$*

At values of $\xi > 1$, the rhomboidal regions dominated by the center-localized distribution in the $S_1$ state merge together across diagonally adjacent unit cells, forming bands aligned in the direction of disrotatory twist which grow progressively wider as $\xi$ increases. Within these bands the $S_1$ adiabatic state is dominated by contributions from the |B> VB state. These regions include the planar configurations. The regions of the $S_1$ state which are dominated by the VB states |L> and |R> are bounded by the $S_1/S_0$ and $S_2/S_1$ intersection pairs, which now all occur along the same direction in the $(\theta_L, \theta_R)$ plane. As the $S_2/S_1$ and $S_1/S_0$ branches of the conical intersection seam approach each other asymptotically in the limit of $\xi$ approaching infinity, these bands will vanish from the $S_1$ state and the population of |B> will be concentrated on $S_1$ for all points in the $(\theta_L, \theta_R)$ plane. Likewise the $S_0$ and $S_2$ states will become dominated by superpositions of |L> and |R> with opposing parity.

### *Transition Dipoles and Photoabsorption Oscillator Strengths*

The intensity of linear absorption from the ground state to an excited state of a system is proportional to the square of the dipole operator sandwiched between the states in question. We have not elaborated on the precise structure of the states of our reaction space, so we cannot make quantitative evaluations of the intensities between the states. However, it is possible to make qualitative statements based upon the symmetry of the states contingent upon the symmetry of a hypothetical molecular frame.

In the context of the geometrical idealization put forward in figure 5, it is natural to consider the case where the untwisted molecular frame possesses $C_{2v}$ symmetry. An examination of the state vectors at untwisted configurations then indicates that in the regime where $\xi < 1$, the $S_1$ state will be a basis for the $B_2$ irreducible representation and the $S_2$ state will be a basis for $A_1$. The dipole moment operator transforms as a $B$ representation, so it can couple the $S_1$ state with either the $S_0$ or $S_2$ states, but cannot couple the latter two with each other. In other words, the $S_1$ state will be bright and the $S_2$ state will be dark in the one-photon absorption spectrum. Conversely, when $\xi > 1$, the reverse ordering holds – $S_1$ is a basis for $A_1$ at planar configurations and $S_2$ is a basis for $B_2$, so that the former is dark while the latter is bright. If the states are close in energy, the intensity may be distributed between them. These observations hold for one-photon absorption. For the case of two-photon absorption, it is quite possible that $S_2$ will be bright when $\xi < 1$. In this case, because $S_1$ can couple to both $S_0$ and $S_2$ via one-photon transitions, it may amplify a two-photon $S_0$-$S_2$ transition via resonant enhancement.

**The Case of Negative $\xi$**

During our exposition of our interaction model, we stated that in the case of negative $\xi$, the model fails to fulfill the basic expectations of a photoisomerization model. The reason for this is that twisting is not favorable on the $S_1$ state in this case. The effect of a negative value of $\xi$ is basically to "flip the model upside-down". This is graphically illustrated in figure 16. Inverting the three potential surfaces means that the identities of the conical intersection seams also switch, though their behavior is analogous to the case where $\xi > 0$. The inversion of the $S_1$ potential energy surface, where twisting is favorable for $\xi > 0$, implies that it is similarly unfavorable for $\xi < 0$. For $\xi < -1$, twisting does become favorable along the disrotatory twist coordinate, where it was not for $\xi > 0$, and leads to an $S_1/S_0$ intersection at the same location in the $(\theta_L, \theta_R)$ plane where an $S_2/S_1$ intersection occurs for $\xi > 1$. It is unfavorable along all alternate directions, however. We conclude that for negative values of the parameter $\xi$, the model does not fulfill the basic expectations of a double bond photoisomerization model.

**DISCUSSION**

*Overview*

We have proposed a model Hamiltonian to describe double bond photoisomerization processes in systems where a π-bond alternation resonance is coupled to transfer of a formal charge. We have based our interaction Hamiltonian on the overlap of fragment orbitals constituting the smallest possible model of such a system – consisting of three orbitals over which a single valence bond resonates between three possible configurations. By the imposition of conditions that can be interpreted as symmetry constraints, we cast the problem in a parameter space to three dimensions, and we have examined the potential surfaces and degeneracy spaces over this parameter space. The reduced dimensionality allows a thorough analysis of the topology and topography of intersection seams over the parameter space. We have offered a detailed interpretation of the parameterization in terms of a hypothetical molecular frame, which illustrated how the reduction of the parameter space can be construed as a particular limit of energetic degeneracy of the basis states and geometric symmetry of the molecular frame. Future works will use this limit as a base from which the model can be expanded so as to provide a more general description of the photoisomerization reactions of methine dyes like those in figure 1. Molecules such as in figure 1 have engendered recent interest as molecular switches in bioimaging applications[27,41,44] or bioelectronic memory arrays[45], and also as model compounds for coherent laser control experiments[30]. Even in its present form, the model provides a physical description that is qualitatively realistic, and even suggests new predictions.

### Core Observations

*Our model fulfills the most basic expectations for a photoisomerization model: 1) it describes favorable bond twisting in excited states 2) it describes low-lying conical intersections associated with bond twist by which the system can decay to the ground state.* The $S_1/S_0$ intersections described by the model can be accessed along a range of twist angles, but occur within a region characterized by the twist of both angles (i.e. along an emergent coordinate).

Our model describes charge localization processes that occur in tandem with bond twisting. It also describes sharp changes in localization in the vicinity of a conical intersection. Charge localization is associated with bond twisting in organic charge-transfer dyes.[70,46] *Ab initio* studies have predicted for such systems that conical intersections join states with different charge localization[43,65,71]. It has been suggested that the charge localized character of twisted states and conical intersections in these reactions may be used as a tool for the design of light-driven molecular devices.[70] Simple photoisomerization models which can capture this behavior may become useful in such pursuits. Twisted intramolecular charge-transfer (TICT) states are a recurring motif seen in *ab initio* studies of photoisomerization in charge-transfer dyes. They are also implicated in more general photoisomerization systems, including those with neutral ground states[34]. However, in charge-transfer dyes – as opposed to polyenes or retinal protonated Schiff base (RPSB) models[23] – it is possible to describe charge localization processes within the manifold of maximally covalent states (fig. 4). This is a key ingredient in our approach, as it allows us to halve the dimensionality of the relevant Hilbert space. The ability to describe charge localization upon bond twisting is an important goal for any photoisomerization model. Our model can describe this behavior.

In addition to being able to describe decay from the first excited state, our model makes interesting suggestions about how higher excited states may become involved in the photoprocesses. Specifically, it suggests that conical intersections between the excited states may occur at planar or near-planar configurations, and describes the induction of barriers on the first excited state surface due to interaction with a higher state. As such, it provides a more detailed electronic structural context to previous phenomenological models of barriers to photoisomerization in methine dyes.[72] It also points to a possible reason why one-photon and two-photon excitation modes lead to fluorescence from a single state in fluorescent protein chromophores[73] and arylmethane dyes[74]. It has been noted for anionic GFP chromophore[73] and for the diarylmethane dye Auramine-O[74] that single- and multi-photon excitation induce fluorescence from the same state. Our model suggests that in systems where $\xi$ is not too close to one, decay from $S_2$ to $S_1$ may require substantial torsion. In restrictive environments, alternate decay pathways may compete effectively and reduce the fluorescence intensity. This is consistent with the pressure-dependence of the fluorescence of Auramine-O in a solid polymer matrix[74] as well as the xanthene dye Rhodamine B[75]. Rhodamine dyes are xanthenes, which are fluorescent dyes engineered through covalent bridging of the rings of arylmethane dyes. Such cross-bridging strategies would be expected to restrict conrotatory motion but may admit hula-twisting motions that are more disrotatory in nature.

### Comparison With Ab Initio Studies of Representative Molecules

There have been *ab initio* calculations on systems analogous to those in figure 1, against which we can compare some of the predictions of our model. Sanchez-Galvez et al. have studied the $S_0$ and $S_1$ surfaces of a complete active space self-consistent field (CASSCF)[76] ansatz for a series of cyanine models[67]. Martin et al.[65], Altoe et al.[63] and Olsen et al.[43,66], have all studied the potential surfaces and wavefunctions of green, red and kindling fluorescent protein chromophore anions with CASSCF methods, and Toniolo et al.[71,77] have calculated minimum energy paths (MEPs) and dynamical trajectories on the $S_1$ surface of a reparameterized semi-empirical complete active space model.

In their extensive survey of the $S_1$ surfaces of a trimethine cyanine model, Sanchez-Galvez et al.[67] noted that the sections of the MEP closest to the Frank-Condon point evolved primarily along conrotatory twist and symmetric stretching coordinates. This result was broadly supported by further results by Improta et al.[64] *Our model successfully describes the initial preference for conrotatory evolution from planar configurations on the $S_1$ state.* Our model is also consistent with symmetric stretching motions in the short-time dynamics, because $\xi$ can be interpreted as encompassing a component of this motion. Our model does not at present accommodate a parameterization of asymmetric stretching. This may explain why our model cannot describe another important observation made by Sanchez-Galvez et al.[67] and Improta et al.[64] for cyanines, or by Martin et al.[65], Altoe et al.[63], and Olsen et al.[43], for fluorescent protein chromophores – the existence of TICT intermediates on the $S_1$ surface. These TICT intermediates are twisted are primarily twisted about a single bond, and they are displaced from $C_2$ symmetry along an antisymmetric bond stretch coordinate. Although our model does predict that twisting of either single bond is favorable in the $S_1$ state, it does not predict minima corresponding to TICT intermediates. Sanchez-Galvez et al.[67] note that the MEP for their cyanine model breaks the initial $C_{2v}$ symmetry in the order $C_{2v} \rightarrow C_2 \rightarrow C_1$. Improta et al.[64] find evidence of similar behavior in their cyanine models. Our analysis of geometrical interpretations for $\xi$ implies that it can only describe totally symmetric modes.

*Our model predicts the existence of $S_1/S_0$ conical intersection seams that span a range of torsion angles.* If one considers the branch of the seam over the entire range of $\xi$, then the seam can be accessed for values of the torsion in the interval $(\pi/2, 3\pi/2)$. Hunt et al.[78] have reported dynamics calculations on the same model cyanine used by Sanchez-Galvez et al.[67] Their results indicate that $S_1/S_0$ intersections can be accessed over the entire range of torsion angles. Sanchez-Galvez et al.[67] reported only low-lying points on the intersection seam for the same model, which were found to lie close to the TICT intermediate structures – in $C_1$ symmetry. The path connecting these minimal-energy conical intersections (MECIs) to the TICT intermediates corresponded to antisymmetric stretching and nitrogen pyramidalization motions. Martin et al.[65] have optimized an $S_0/S_1$ conical intersection point which displays disrotatory ('hula') twist. Our model is not consistent with this, and only predicts disrotatory-twisted $S_0/S_1$ CIs at the 3-state intersection. Our model does not have the dimensionality required to span the branching space of a 3-state intersection[79]. The branching space of a three-state intersection is, in general, five-dimensional, in line with the requirements of zero diagonal splitting and zero coupling. This five-dimensional branching space contains two three-dimensional two-state intersection seams[79]. The failure of the model to predict disrotatory-twisted $S_0/S_1$ CIs may arise from the restricted dimensionality of our parameterization. Olsen et al.[43] and Toniolo et al.[71] have reported low-lying intersection points nearby to TICT intermediates for a model of GFP chromophore anion. These minimal energy conical intersection points showed an uneven distribution of twist on the bridge, with most of the twist concentrated on a single bond. Our model does not describe these intersections because it lacks a suitable parameterization for antisymmetric stretching motions. To our knowledge, low-lying intersections along conrotatory twist coordinates have not been reported for similar systems. This may be because characterization of conical intersection seams is currently dominated by discussion of minimal energy points on the seams. If a portion of intersection seam with conrotatory twist ($C_{2v}$ symmetric) were to be found for a symmetric charge-transfer dye, than this may be considered support for the current model.

Pyramidalization or re-hybridization of carbon centers is commonly observed in minimum-energy geometries of charge-transfer conical intersection seams[34]. We are considering generalizations of our model that can describe this motion. In the context of GFP chromophore anions, the geometries of minimal energy conical intersections that have been reported all feature visible pyramidalization of the methine bridge carbon[43,65]. This is not the case, however, for models of kindling (KFP) or red (RFP) fluorescent protein chromophores[43,66]. In these latter cases, the intermediates and MECIs corresponding to twist about different bonds are not degenerate on the $S_1$ surface, and only the higher-energy MECI features pyramidalization. Pyramidalization motions may be important because they could stabilize the ionic state manifold and thereby call into question the assumptions built into our ansatz for the low-lying states (Fig.4 ). Analysis of CASVB[80] wavefunctions near the MECIs which we have calculated for FP chromophore anions suggest that this is not the case.[43]

Our model in its current form only considers interactions between degenerate valence bond states. We have not, at this time, attempted to parameterize the diagonal splitting which will arise between the states for a realistic molecule. This may well be the underlying reason why our model does not describe TICT intermediates on the excited state surface. The inclusion of diagonal biasing may also manifest itself in the distribution of conical intersection points in the model. The multidimensionality of the branching space of conical intersections arises from the need to simultaneously zero both the diagonal splitting and the off-diagonal coupling of an effective Hamiltonian. For a two-state intersection, this requires two degrees of freedom; in the three-state case it requires five. By concentrating only on the state interactions, we have, in effect, constrained our system to a three-dimensional cross-section through the intrinsically five-dimensional space of 3 x 3 symmetric matrices. The intersection seam visualized in figure 14 represents the union of two one-dimensional projections of three-dimensional manifolds representing the two-state intersection seams. One may expect that introducing energetic bias amongst the states will alter the position of the intersection points projected onto the three-dimensional space that we have used to parameterize our model. This may result in the appearance of additional sections of intersection seam at accessible energies – for example allowing decay at values of the twist angles that our model does not currently predict. However, as long as the parameterization of the matrix elements is continuous and single-valued modulo any periodic boundary conditions then this will also be true for the intersection seam. We will deal with extensions of our model that include state biasing – and the parameterization of this biasing – in a future publication.

### *Prospects*

Our current model has the potential for use in studies of the photoisomerization of coupled double bonds and charge-transfer dyes. It is suitable in extensions of empirical valence force fields, in a similar spirit to other techniques which couple molecular mechanics force fields to valence-bond-like Hamiltonians[14,16]. One of the key problems with single-state molecular mechanics force fields in addressing the properties of systems in figure 1 is that they do not have the scope to describe the change from resonating planar structures to highly polar twisted intramolecular charge-transfer (TICT) states. We feel, for example, that this is a key problem in the invocation of molecular mechanical results for the ground state isomerization of the chromophore in fluorescent proteins to interpret the excited state isomerization processes[60,81]. Charge localization is a common motif seen in *ab initio* treatments of photoisomerization reactions[70]. Our work suggests that this qualitative behavior could be described for methine dyes using a parameterized interaction model defined over a covalent electronic Hilbert space of reasonably low dimension.

There are certain features of photoisomerization in methine dyes that our model cannot describe in its current form. These include TICT intermediates on the $S_1$ surface, as well $S_1/S_0$ conical intersection seams at geometries with uneven bridge twist or with significant disrotatory twist. Our model is easily generalizable, and we expect that many of these features become accessible as the model is expanded. As this occurs, it will illuminate further the underlying physics of methine photoisomerization. We are currently investigating appropriate generalizations and will describe these in upcoming papers.

The multi-dimensional nature of molecular decay processes presents formidable difficulty when one attempts to understand the fundamental physics. This is true for molecules in isolation, but the difficulty becomes even more acute when one wishes to describe environmental influences or quantum dynamical effects. Even in the case of isolated ethylene, the simplest double bonding system, consideration of the full intersection space requires ten dimensions. In this context, reduced models are useful because they provide fundamental insight as well as a point of reference from which extensions can be made in an informed and systematic manner. They have been used in this manner to great effect in understanding the photoisomerization processes of ethylenic systems.[19,23] In this sense, our model carries considerable pedagogical value in its current form. By virtue of its restricted dimensionality, we have been able to clearly visualize the conical intersection seams which lead to excited-state decay. Further extensions of our interaction Hamiltonian will rely heavily on the current results as a point of reference for understanding how the structure of the intersections space evolves as the dimensionality increases.

ACKNOWLEDGMENT This work was partially supported with funds from the Australian Research Council Discovery Project DP0877875. We thank Anthony Jacko and Michael Smith for readings of the manuscript. We additionally thank Anthony Jacko for bringing to our attention an error in the eigenvalues formulas. Some of the graphics were generated with the Mathematica[82], VMD[83] and ChemBioDraw[58] software packages.

FIGURE CAPTIONS

**Figure 1.** Examples of charge-transfer dyes which display double bond isomerization photochemistry. All three molecules resonate between Lewis structures which feature invert bond alternation and relocates the formal charge. (Top) The chromophore of the green fluorescent protein (GFP) in its anionic form. The cyanine dye Thiazole Orange (middle) and the triarylmethane dye Malachite Green (bottom) become fluorescent when bound to biological macromolecules, making them useful biotechnological stains.

**Figure 2.** The 4 isomers of the asymmetrical dye Thiazole Orange, which differ by (Z,E) isomerism of the bridge. Thiazole Orange is the least symmetrical of the three example molecules in Fig. 1, and so all isomers are distinguishable. They are labelled according to standard organic chemistry nomenclature.

**Figure 3.** Important concepts in photochemical mechanisms. Reactants (green lump) are promoted to the excited state surface by a photon, forming the Frank Condon State (red lump) which then evolves in in all possible ways (transparent orange lumps along black lines) on the excited state. Upon passing near a conical intersection between the surfaces, population can return to the ground state and continues to evolve to form one or more products (yellow lumps).

**Figure 4.** Electronic states of a model charge-transfer dye. States are generated by arranging four electrons (excess positive charge) or two electrons (excess negative charge) amongst $\pi$ orbitals on different fragments. Localized complete active space orbitals for a green fluorescent protein chromophore model are shown as an example. The 'nuclei' are assumed to carry a charge of +1, so single occupation neutralizes the 'atom'. For the four-electron case, three maximally covalent configurations and three ionic configurations are shown. The ionic configurations are generated by separating the charge in the bond of one of the covalent structures. The covalent and ionic state spaces can be contracted to form three "perfect-pairing" valence bond wavefunctions that are distinguishable by the location of the bond and the formal charge.

**Figure 5.** (Top) The parameterization of our interaction Hamiltonian. The off-diagonal elements are listed next to the line connecting the sites that they couple. The coupling of the end sites to the bridge are given by cosine terms representing π pond overlap. The third interaction element is the cosine of the difference between bridge-end angles multiplied by a scaling parameter. (Bottom) A geometrical idealization of the model. The angles $\theta_L$ and $\theta_R$ have a clear geometrical interpretation as the torsion angles of the bonds connecting the end sites to the bridge. The parameter $\xi$ represents a scaling of end-end relative to end-bridge coupling.

**Figure 6.** A geometric interpretation of the parameter $\xi$ of our model. Contours of constant $\xi$ are plotted against the internal angle $\phi$ and the quantity $\rho$, which is directly proportional to the length of the bridge-end bond, in the context of the geometrical model presented in figure 6. The parameter $\xi$ maps to curved lines in the plane, and can be thought of as a combination of angle contraction and symmetric bond stretching in this context. The parameters appropriate to an allylic anion would occur near $(\rho, \phi) = (4.37, 2\pi/3)$.

**Figure 7.** (Top) The eigenvalues of a traceless 3 x 3 matrix over the $(u,v)$ plane. The quantities $u$ and $v$ (eqn. 7 & 8) completely specify the energy splittings of a 3 x 3 symmetric and positive definite matrix. The conditions $u < 0$ and $u^3 + v^2 < 0$ (region shown at bottom) are sufficient to guarantee real eigenvalues of a 3 x 3 matrix and are equivalent to symmetry and positive definiteness of the matrix. When the inequality is strong, 3 non-degenerate eigenvalues exist. On the boundary (highlighted in black) at least 2 of the eigenvalues are degenerate. $S_1/S_0$ degeneracies occur on the $v < 0$ part of the boundary. $S_2/S_1$ degeneracies occur on the $v > 0$ region of the boundary, and a 3-state intersection occurs at $(u,v) = (0,0)$.

**Figure 8.** Schematic representation of relevant coordinates in terms of the geometrical model in figure 6 (top) and their representation as vectors in the $(\theta_L, \theta_R)$ plane (bottom). Single bond twists change one angle while leaving the other constant; they represent the torsion of one of the two π bonds. The conrotatory and disrotatory twist coordinates are combinations of the single bond twists, respectively.

They are emergent coordinates with no clear analogue for a subsystem composed of a single π bond. In the context of an untwisted molecular frame with $C_{2v}$ symmetry, the conrotatory twist preserves $C_2$ symmetry and breaks $C_s$ symmetry, while the disrotatory twist breaks $C_s$ symmetry and preserves $C_2$ symmetry.

**Figure 9.** Potential surfaces of the model when $\xi = 0$. A three-dimensional view is shown at the top of the figure. Colored dashed lines through the surfaces represent one-bond (green), disrotatory (blue) and conrotatory (red) twisting motions. Profiles of the surfaces along these slices are shown at bottom, in the same color scheme.

**Figure 10.** Potential surfaces of the model when $\xi = 0.5$, representing the situation arising in the regime $0 < \xi < 1$. A three-dimensional view is shown at the top of the figure. Colored dashed lines through the surfaces represent one-bond (green), disrotatory (blue) and conrotatory (red) twisting motions. Profiles of the surfaces along these slices are shown at bottom, in the same color scheme. Within this regime of the model the $S_2/S_1$ and $S_1/S_0$ conical intersections occur along orthogonal slices. The $S_2/S_1$ intersections occur along the disrotatory twist slice (blue). The $S_1/S_0$ intersections occur along the conrotatory twist slice (red). Bond twisting motions are generally favorable on both excited states, and unfavorable on the ground state.

**Figure 11.** Potential surfaces of the model when $\xi = 1.0$, representing the situation arising at the critical level $\xi = 1$. A three-dimensional view is shown at the top of the figure. Colored dashed lines through the surfaces represent one-bond (green), disrotatory (blue) and conrotatory (red) twisting motions. Profiles of the surfaces along these slices are shown at bottom, in the same color scheme. At this critical level of $\xi$, the $S_2/S_1$ intersections occur at planar configurations ($\theta_{L,R} = 0 + n\pi$). They are no longer conical intersections, but are glancing (Renner-Teller) intersections instead. The $S_1/S_0$ intersections occur along the conrotatory twist slice (red) as the $0 < \xi < 1$ regime. Bond twisting motions are generally favorable on the first excited state and unfavorable on the ground state, from planar configurations. The highest state ($S_2$) is flat with respect to all bond-twisting motions.

**Figure 12.** Potential surfaces of the model when $\xi = 1.5$, representing the regime $1 < \xi$. A three-dimensional view is shown at the top of the figure. Colored dashed lines through the surfaces represent one-bond (green), disrotatory (blue) and conrotatory (red) twisting motions. Profiles of the surfaces along these slices are shown at bottom, in the same color scheme. Within this regime of the model the $S_2/S_1$ and $S_1/S_0$ conical intersections occur along the line representing pure conrotatory twist motion. Bond twisting motions on the first excited state ($S_1$) are now activated along the conrotatory twist direction, and lead to the $S_2/S_1$ intersection. Disrotatory twisting is favorable on the $S_1$ state. Bond twisting is favorable along one-bond and conrotatory twisting directions in the upper state ($S_2$). Pure conrotatory twist motion on this surface leads to the $S_2/S_1$ intersection. Bond twisting on the ground state is unfavorable from planar configurations.

**Figure 13.** Eigensurfaces over a plane spanned by $\xi$ and conrotatory twist (left top), the disrotatory twist (right top), and one-bond twists (bottom) obtained by twisting one bond while the other is untwisted (bottom left) or while the other is twisted at an angle of $\pi/2$ (bottom right).

**Figure 14.** The conical intersection seam over the 3-dimensional parameter space. The intersections in a unit cell of the $(\theta_L, \theta_R)$ plane (left) are continued over the periodic boundaries as the cell is extended (top right). Contours are given by the equation $|v| - |u| = 0.005$ (see eqns. 15 & 16) and forms a tubular contour enclosing the intersection seam. Contours are colored according to the sign of the angle $\Theta$ (see eq. 17); red contours enclose $S_1/S_0$ intersections and yellow contours enclose $S_2/S_1$ intersections. The $S_2/S_1$ intersection seams from diagonally adjacent unit cells collide and diverge at the corners, changing direction when $\xi = 1.0$. The identity of the intersecting pairs of states is simply related to the sign of the parameter $v$, which determines whether the intersection corresponds to $\Theta = 0$ ($S_2/S_1$ intersection) or $\Theta = \pi$ ($S_1/S_0$ intersection) at points along the seam. At $\xi = 0.0$, the model becomes singular at $(\pi/2, \pi/2)$, giving rise to a 3 state intersection and an indeterminate value of $\Theta$. Variations in color on the contour near this point reflect increasing rapid variations in $\Theta$. As $\xi$ becomes infinite, the $S_2/S_1$ and $S_1/S_0$ intersections will asymptotically approach each other at the points $(\pi/4, \pi/4)$ and $(3\pi/4, 3\pi/4)$.

**Figure 15.** Populations of the VB basis states over the $(\theta_L, \theta_R)$ plane for adiabatic states $S_0$ (bottom), $S_1$ (middle) and $S_2$ (top) for $\xi$ values (left to right) 0.0, 0.5, 1.0 and 1.5. The basis states have been mapped to the RGB color system using the legend shown at top. The color at any one point was generated by using the squared components of the adiabatic state vectors as RGB coordinates. Fast changes in the wavefunctions are observed in the vicinity of all conical intersections, but are most dramatic in the vicinity of the conical intersections, where the states change character discontinuously.

**Figure 16.** Features of the model when $\xi$ is allowed to become negative. The results obtained for our model when $\xi < 0$ are easily extrapolated from results obtained for $\xi > 0$. The results can be obtained by "turning the model on its head", in the sense that the shifted eigenvalues of the model over the $(\theta_L, \theta_R)$ plane are mirrored through $z = 0$ (top of figure). As a consequence of this behavior, the identities of the intersections change (bottom), with $S_1/S_0$ intersections becoming $S_2/S_1$ intersections, etc. Twisting is generally not favorable on the $S_1$ surface for $\xi < 0$, so the model in this regime does not fulfill the basic expectations of a double bond photoisomerization model.

## REFERENCES

(1)     Malrieu, J.-P., *Theo. Chem. Acc.* **1981**, *59*, 251-279.
(2)     Malrieu, J. P.; Nebot-Gil, I.; Sánchez-Marĺn, J., *Pure And Applied Chemistry* **1984**, *56*, 1241-1254.
(3)     Boggio-Pasqua, M.; Bearpark, M. J.; Klene, M.; Robb, M. A., *J. Chem. Phys.* **2004**, *120*, 7849-7860.
(4)     Miller, W. H., *Proc. Natl. Acad. Sci. USA* **2005**, *102*, 6660-6664.
(5)     Helms, V., *Curr. Op. Struct. Biol.* **2002**, *12*, 169-175.
(6)     LaVan, D. A.; Cha, J. N., *PNAS* **2006**, *103*, 5251-5255.
(7)     Schreiber, M.; Silva-Junior, M. R.; Sauer, S. P. A.; Thiel, W., *J. Chem. Phys.* **2008**, *128*, 134110-25.
(8)     Warshel, A.; Levitt, M., *J. Mol. Bio* **1976**, *103*, 227-249.
(9)     Klähn, M.; Braun-Sand, S.; Rosta, E.; Warshel, A., *J. Phys. Chem.* **2005**, *B109*, 15645-15650.
(10)    Reuter, N.; Dejaegere, A.; Maigret, B.; Karplus, M., *J. Phys. Chem.* **2000**, *A104*, 1720-1735.
(11)    Ben-Nun, M.; Martinez, T. J., *Chem. Phys. Lett.* **1998**, *290*, 289-295.
(12)    Bianco, R.; Timoneda, J. J. i.; Hynes, J. T., *J. Phys. Chem.* **1994**, *47*, 12103-12107.
(13)    Malrieu, J.-P., *J. Mol. Struct. (THEOCHEM)* **1998**, *424*, 83-91.
(14)    Warshel, A.; Weiss, R. M., *J. Am. Chem. Soc.* **1980**, *102*, 6218-6226.

(15)     Ben-Nun, M.; Molnar, F.; Lu, H.; Phillips, J. C.; J., T.; Martĺnez; Schulten, K., *Faraday Discussions* **1988**, *110*, 447-462.

(16)     Treboux, G.; Maynau, D.; Malrieu, J. P., *J. Phys. Chem.* **1995**, *99*, 6417-6423.

(17)     Sason Shaik, A. S., *Ang. Chem. Int. Ed.* **1999**, *38*, 586-625.

(18)     Voth, G. A., *Acc. Chem. Res.* **2006**, *39*, 143-150.

(19)     Burghardt, I.; Hynes, J. T., *J. Phys. Chem.* **2006**, *A110*, 11411-11423.

(20)     Thompson, W. H.; Blanchard-Desce, M.; Hynes, J. T., *J. Phys. Chem.* **1998**, *A102*, 7712-7722.

(21)     Thompson, W. H.; Blanchard-Desce, M.; Alain, V.; Muller, J.; Fort, A.; Barzoukas, M.; Hynes, J. T., *J. Phys. Chem.* **1999**, *A103*, 3766-3771.

(22)     Laage, D.; Thompson, W. H.; Blanchard-Desce, M.; Hynes, J. T., *J. Phys. Chem.* **2003**, *A107*, 6032-6046.

(23)     Bonačić-Koutecký, V.; Koutecký, J.; Michl, J., *Ang. Chem. Int. Ed.* **1987**, *26*, 170-189.

(24)     Shaik, S., *New Journal of Chemistry* **2007**, *31*, 2015-2028.

(25)     Berneth, H. In *Ullmann's Encyclopedia of Industrial Chemistry*; Wiley-VCH Verlag GmbH & Co. KGaA.: 2008.

(26)     Dugave, C.; Demange, L., *Chem. Rev.* **2003**, *103*, 2475-2532.

(27)     Furstenberg, A.; Julliard, M. D.; Deligeorgiev, T. G.; Gadjev, N. I.; Vasilev, A. A.; Vauthey, E., *J. Am. Chem. Soc.* **2006**, *128*, 7661-7669.

(28)     Chudakov, D. M.; Lukyanov, S.; Lukyanov, K. A., *Trends in Biotechnology* **2005**, *23*, 605-613.

(29)     Babendure, J. R.; Adams, S. R.; Tsien, R. Y., *J. Am. Chem. Soc.* **2003**, *125*, 14716-14717.

(30)     Vogt, G.; Krampert, G.; Niklaus, P.; Nuernberger, P.; Gerber, G., *Phys. Rev. Lett.* **2005**, *94*, 068305-4.

(31)     Worth, G. A.; Cederbaum, L. S., *Ann. Rev. Phys. Chem.* **2004**, *55*, 127-158.

(32)     Yarkony, D. R., *Rev. Mod. Phys.* **1996**, *68*, 985 LP  - 1013.

(33)     Yarkony, D. R., *J. Phys. Chem.* **2001**, *A105*, 6277-6293.

(34)     Levine, B. G.; Martinez, T. J., *Ann. Rev. Phys. Chem.* **2007**, *58*, 613-634.

(35)     Levine, B. G.; Coe, J. D.; Martinez, T. J., *J. Phys. Chem.* **2008**, *B112*, 405-413.

(36)     Sicilia, F.; Blancafort, L.; Bearpark, M. J.; Robb, M. A., *J. Chem. Theory Comp.* **2008**, *4*, 257-266.

(37)     Ko, C.; Levine, B.; Toniolo, A.; Manohar, L.; Olsen, S.; Werner, H.-J.; Martinez, T. J., *J. Am. Chem. Soc.* **2003**, *125*, 12710-12711.

(38)     Sicilia, F.; Bearpark, M.; Blancafort, L.; Robb, M., *Theo. Chem. Acc.* **2007**, *118*, 241-251.

(39)     Hines, A. P.; Dawson, C. M.; McKenzie, R. H.; Milburn, G. J., *Phys. Rev.* **2004**, *A70*, 022303.

(40)     Jaeger, G., *Quantum Information: An Overview*; 1 ed.; Springer: New York, 2006. p. 38.

(41)     Netzel, T. L.; Nafisi, K.; Zhao, M.; Lenhard, J. R.; Johnson, I., *J. Phys. Chem.* **1995**, *99*, 17936-17947.

(42)     Kummer, A.; Kompa, C.; Lossau, H.; Pöllinger-Dammer, F.; Michel-Beyerle, M.; Silva, C. M.; Bylina, E. J.; Coleman, W. J.; Yang, M. M.; Youvan, D. C., *Chem. Phys.* **1998**, *237*, 183-193.

(43)     Olsen, S.; Smith, S. C., *J. Am. Chem. Soc.* **2008**, *130*, 8677-8689.

(44)     Silva, G. L.; Ediz, V.; Yaron, D.; Armitage, B. A., *J. Am. Chem. Soc.* **2007**, *129*, 5710-5718.

(45)     Sauer, M., *Proc. Natl. Acad. Sci. USA* **2005**, *102*, 9433-9434.

(46)     Henderson, J. N.; Remington, S. J., *Physiology* **2006**, *21*, 162-170.

(47)     Shaik, S.; Hiberty, P. C., *A Chemist's Guide to Valence-Bond Theory*; Wiley, 2008

(48)     Shaik, S.; Philippe C. Hiberty In *Reviews in Computational Chemistry*; Lipkowitz, K. B., Larter, R., Cundari, T. R., Eds.; VCH Publishers: New York, 2004; Vol. 20, p 1-100.

(49)     Pacher, T.; Cederbaum, L. S.; Koppel, H., *J. Chem. Phys.* **1988**, *89*, 7367-7381.
(50)     McWeeny, R., *Proc. Roy. Soc. London* **1954**, *A223*, 306-323.
(51)     Foster, J. M.; Boys, S. F., *Rev. Mod. Phys.* **1960**, *32*, 300.
(52)     Stalring, J.; Bernhardsson, A.; Lindh, R., *Mol. Phys.* **2001**, *99*, 103-114.
(53)     Cederbaum, L. S.; Schirmer, J.; Meyer, H. D., *J. Phys.* **1989**, *A22*, 2427.
(54)     Mulliken, R. S.; Rieke, C. A.; Orloff, D.; Orloff, H., *J. Chem. Phys.* **1949**, *17*, 1248-1267.
(55)     Tolbert, L. M., *Acc. Chem. Res.* **1986**, *19*, 268-273.
(56)     Tolbert, L. M.; Ali, M. Z., *J. Org. Chem.* **1985**, *50*, 3288-3295.
(57)     Cocolicchio, D.; Viggiano, M., *J. Phys.* **2000**, *A33*, 5669-5673.
(58)     CambridgeSoft, I.; 11.0 ed. Cambridge, MA, 2008.
(59)     Liu, R. S. H., *Acc. Chem. Res.* **2001**, *34*, 555-562.
(60)     Baffour-Awuah, N. Y. A.; Zimmer, M., *Chem. Phys.* **2004**, *303*, 7-11.
(61)     Wallace, A., *Differential Topology: First Steps*; Dover Publications Inc.: Mineola, N.Y., 2006. p. 10.
(62)     Vértesi, T.; Bene, E., *Chem. Phys. Lett.* **2004**, *392*, 17-22.
(63)     Altoe, P.; Bernardi, F.; Garavelli, M.; Orlandi, G.; Negri, F., *J. Am. Chem. Soc.* **2005**, *127*, 3952-3963.
(64)     Improta, R.; Santoro, F., *J. Chem. Theory Comp.* **2005**, *1*, 215-229.
(65)     Martin, M. E.; Negri, F.; Olivucci, M., *J. Am. Chem. Soc.* **2004**, *126*, 5452-5464.
(66)     Olsen, S.; Smith, S. C., *J. Am. Chem. Soc.* **2007**, *129*, 2054-2065.
(67)     Sanchez-Galvez, A.; Hunt, P.; Robb, M. A.; Olivucci, M.; Vreven, T.; Schlegel, H. B., *J. Am. Chem. Soc.* **2000**, *122*, 2911-2924.
(68)     Zilberg, S.; Haas, Y., *J. Phys. Chem.* **1998**, *A102*, 10843-10850.
(69)     Zilberg, S.; Haas, Y., *J. Phys. Chem.* **1999**, *A103*, 2364-2374.
(70)     Martínez, T. J., *Acc. Chem. Res.* **2006**, *39*, 119-126.
(71)     Toniolo, A.; Olsen, S.; Manohar, L.; Martinez, T. J., *Faraday Discussions* **2004**, *129*, 149-163.
(72)     Rullière, C., *Chem. Phys. Lett.* **1976**, *43*, 303-308.
(73)     Hosoi, H.; Yamaguchi, S.; Mizuno, H.; Miyawaki, A.; Tahara, T., *J. Phys. Chem.* **2008**, *B112*, 2761-2763.
(74)     Dreger, Z. A.; Yang, G.; White, J. O.; Li, Y.; Drickamer, H. G., *J. Phys. Chem.* **1997**, *A101*, 9511-9519.
(75)     Dreger, Z. A.; Yang, G.; White, J. O.; Li, Y.; Drickamer, H. G., *J. Phys. Chem.* **1998**, *B102*, 4380-4385.
(76)     Roos, B. O. In *Ab Initio Methods in Quantum Chemistry II*; Lawley, K. P., Ed.; John Wiley and Sons: New York, 1987, p 399.
(77)     Toniolo, A.; Olsen, S.; Manohar, L.; Martínez, T. J. In *Femtochemistry and Femtobiology: Ultrafast Events in Molecular Science*; M. Martin, Hynes, J. T., Eds.; Elsevier: Amsterdam, 2004.
(78)     Hunt, P. A.; Robb, M. A., *J. Am. Chem. Soc.* **2005**, *127*, 5720-5726.
(79)     Schuurman, M. S.; Yarkony, D. R., *J. Chem. Phys.* **2006**, *124*, 244103-11.
(80)     Hirao, K.; Nakano, H.; Nakayama, K.; Dupuis, M., *J. Chem. Phys.* **1996**, *105*, 9227-9239.
(81)     Moors, S. L. C.; Michielssens, S.; Flors, C.; Dedecker, P.; Hofkens, J.; Ceulemans, A., *J. Chem. Theory Comp.* **2008**, *4*, 1012-1020.
(82)     Wolfram Research, I.; 6.0 ed. Champaign, IL, 2008.
(83)     Humphrey, W.; Dalke, A.; Schulten, K., *Journal of Molecular Graphics* **1996**, *14*.

*GFP Chromophore Anion*

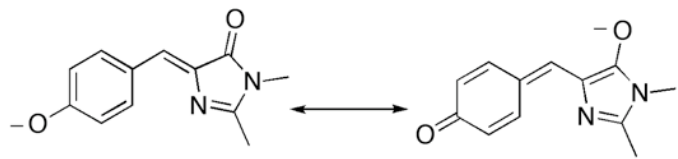

*Thiazole Orange*

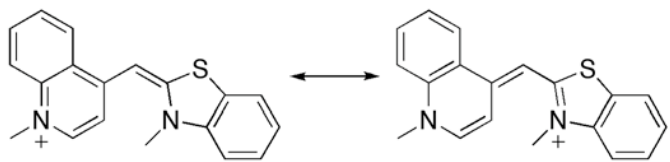

*Malachite Green*

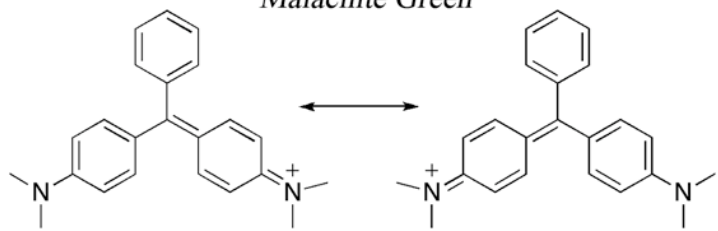

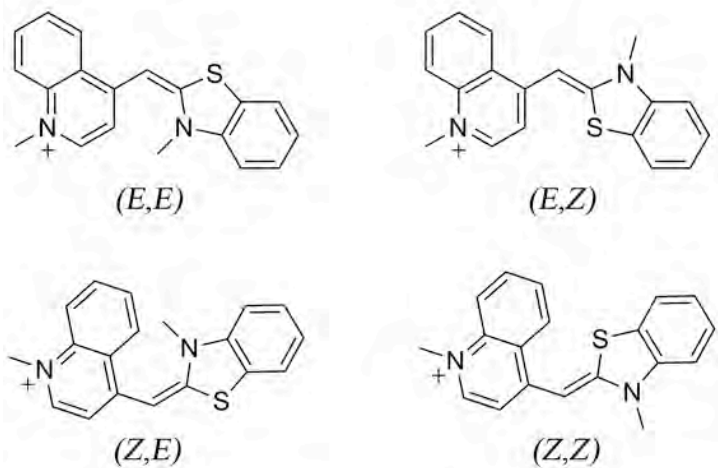

(E,E)                    (E,Z)

(Z,E)                    (Z,Z)

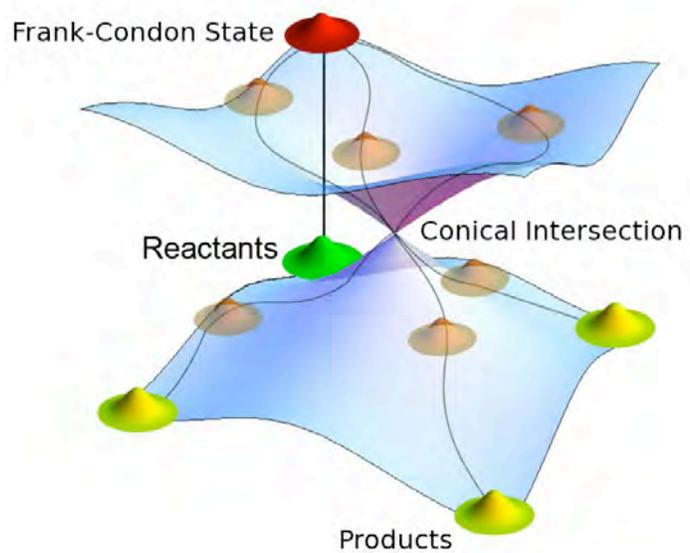

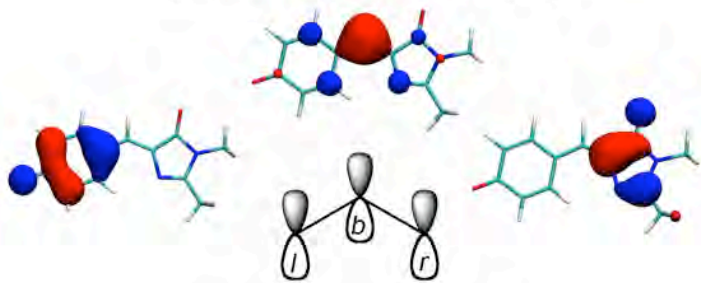

Fragment π Orbitals
*(GFP Chromophore as example)*

## Covalent Structures

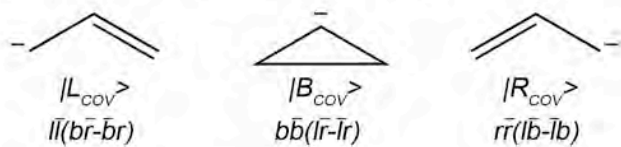

## Ionic Structures

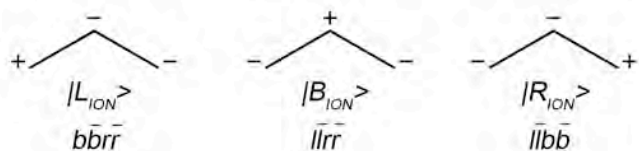

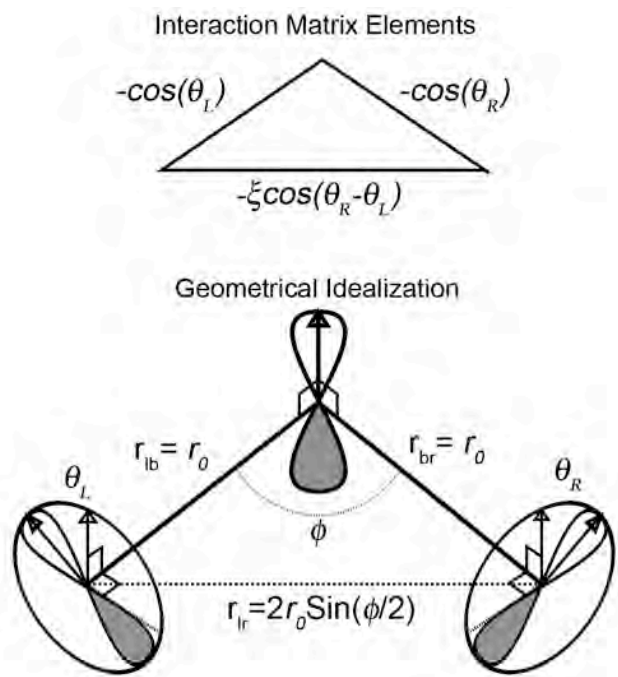

## Geometric Interpretation of $\xi$

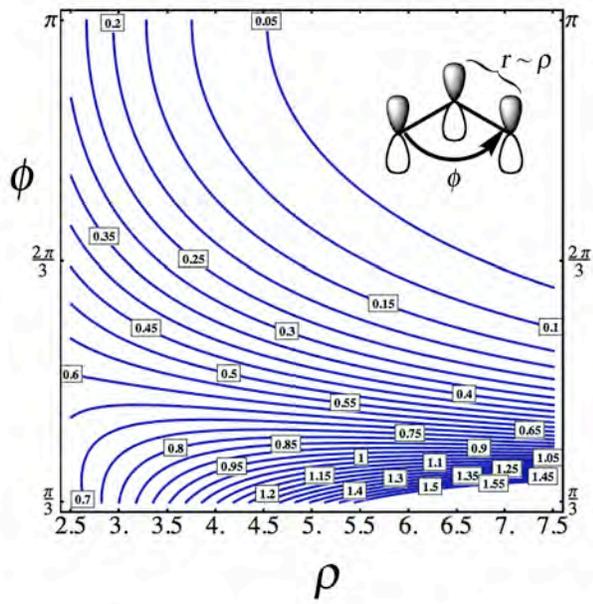

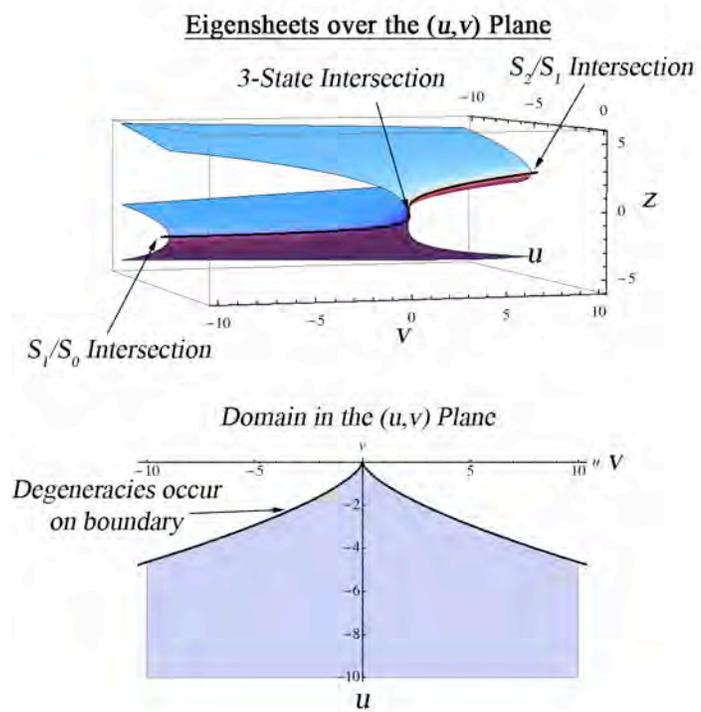

**Eigensheets over the (u,v) Plane**

*3-State Intersection*

$S_2/S_1$ Intersection

$S_1/S_0$ Intersection

z

u

V

Domain in the (u,v) Plane

*Degeneracies occur on boundary*

v

u

V

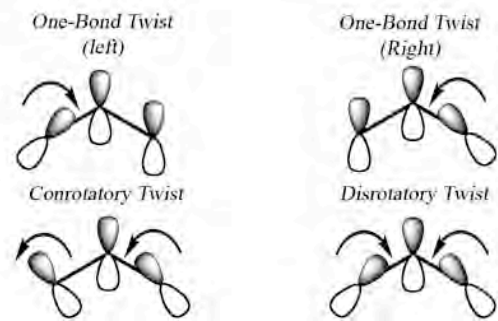

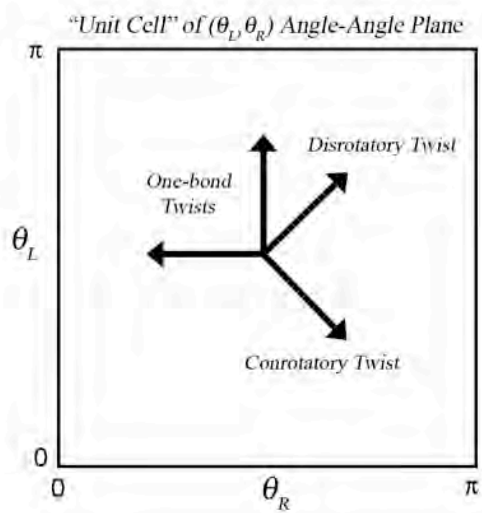

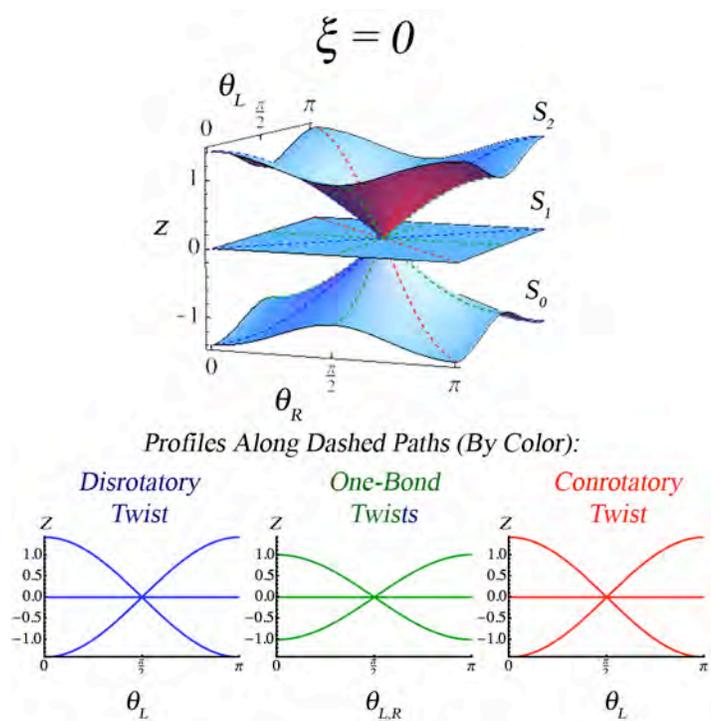

$$\xi = 0$$

Profiles Along Dashed Paths (By Color):

Disrotatory Twist | One-Bond Twists | Conrotatory Twist

$$\xi = 0.5$$

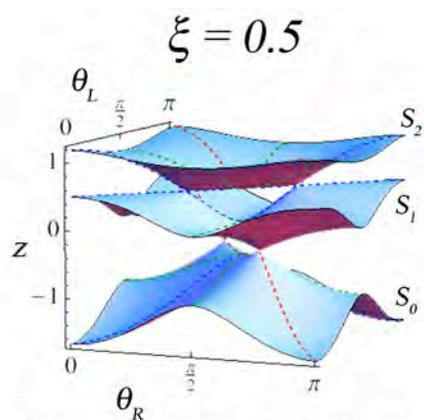

Profiles Along Dashed Paths (By Color):

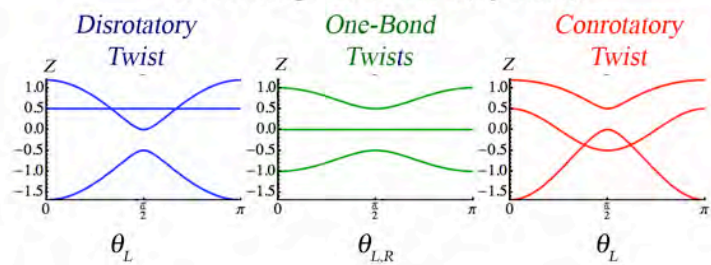

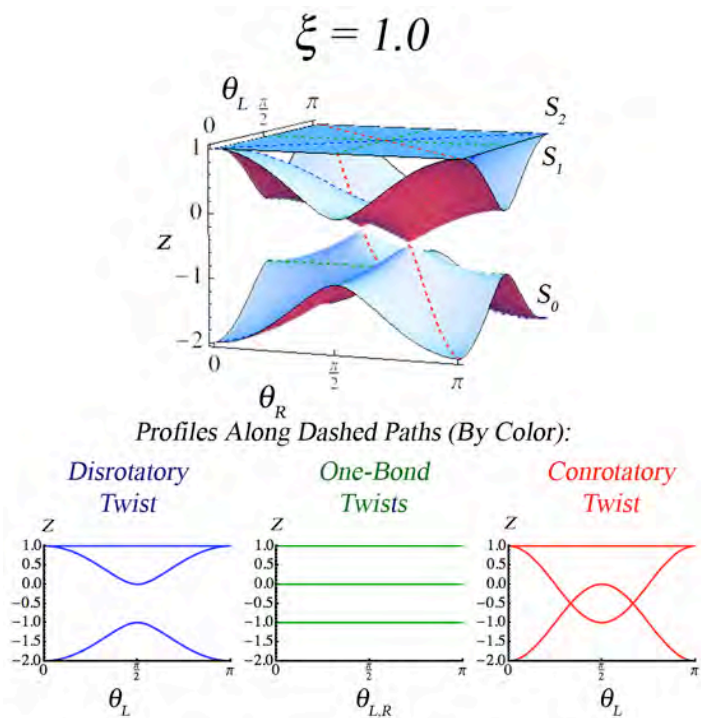

## ξ = 1.0

*Profiles Along Dashed Paths (By Color):*

| Disrotatory Twist | One-Bond Twists | Conrotatory Twist |
|---|---|---|

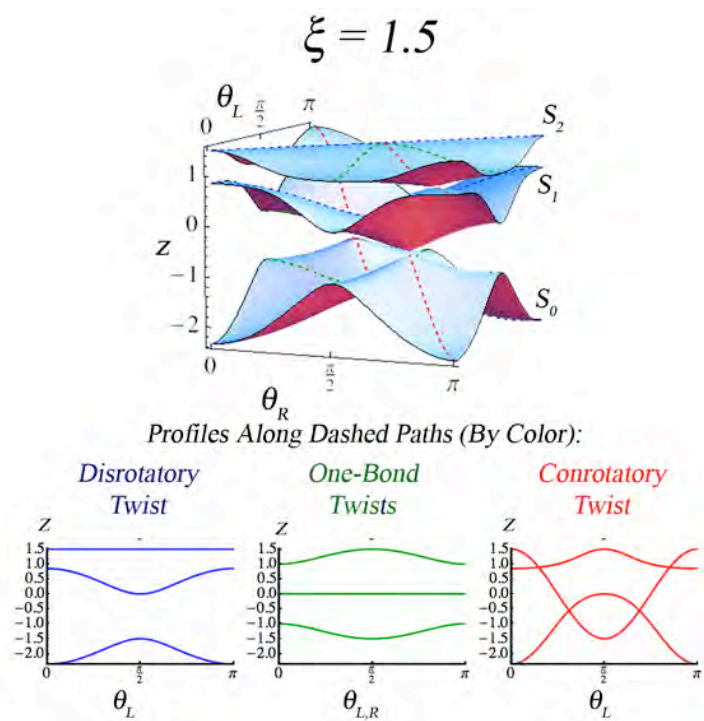

$\xi = 1.5$

Profiles Along Dashed Paths (By Color):

Disrotatory Twist     One-Bond Twists     Conrotatory Twist

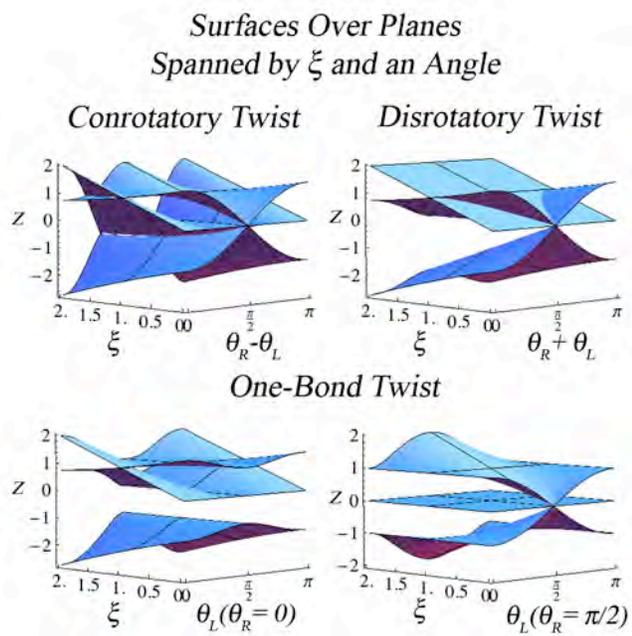

*Surfaces Over Planes*
*Spanned by ξ and an Angle*

*Conrotatory Twist*     *Disrotatory Twist*

*One-Bond Twist*

Figure 14

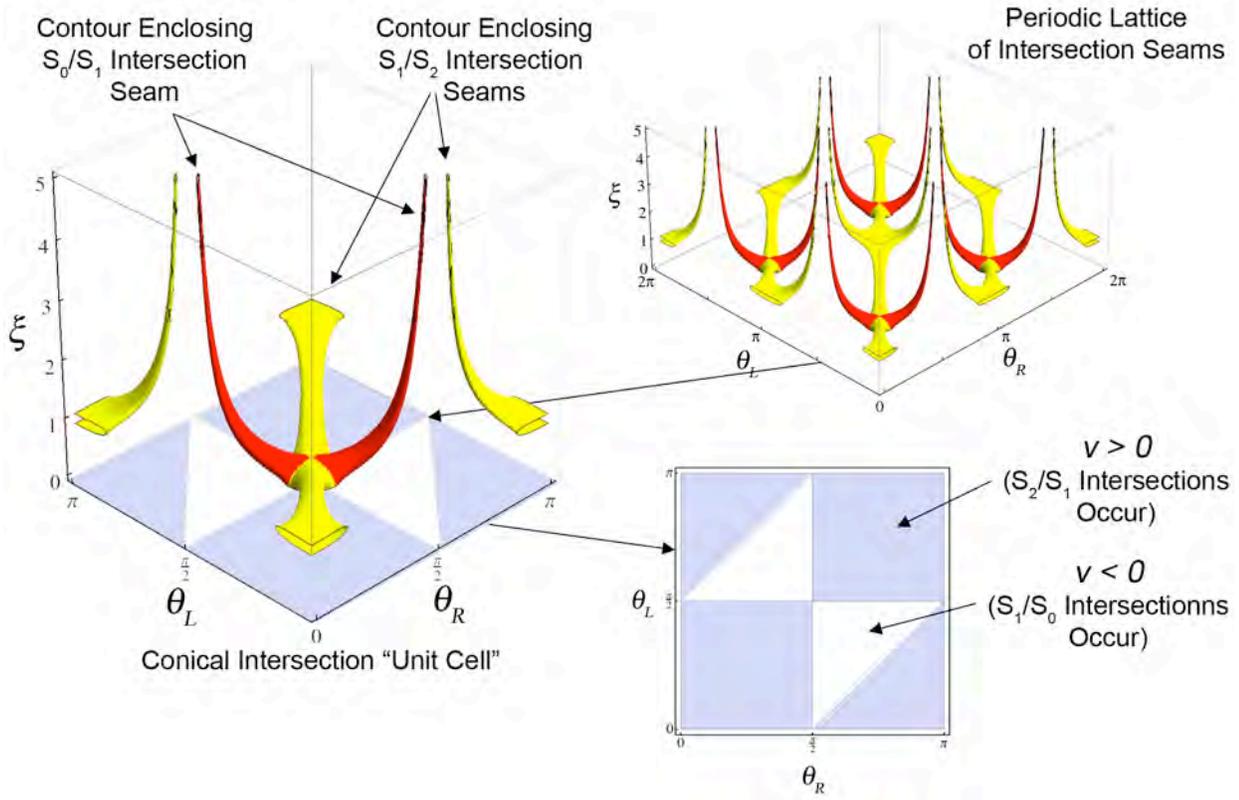

Contour Enclosing
$S_0/S_1$ Intersection
Seam

Contour Enclosing
$S_1/S_2$ Intersection
Seams

Periodic Lattice
of Intersection Seams

$\xi$

$\theta_L$

$\theta_R$

Conical Intersection "Unit Cell"

$v > 0$
($S_2/S_1$ Intersections
Occur)

$v < 0$
($S_1/S_0$ Intersectionns
Occur)

$\theta_L$

$\theta_R$

## Populations of Valence-Bond Basis States

*RGB Color Map:*

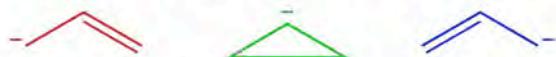

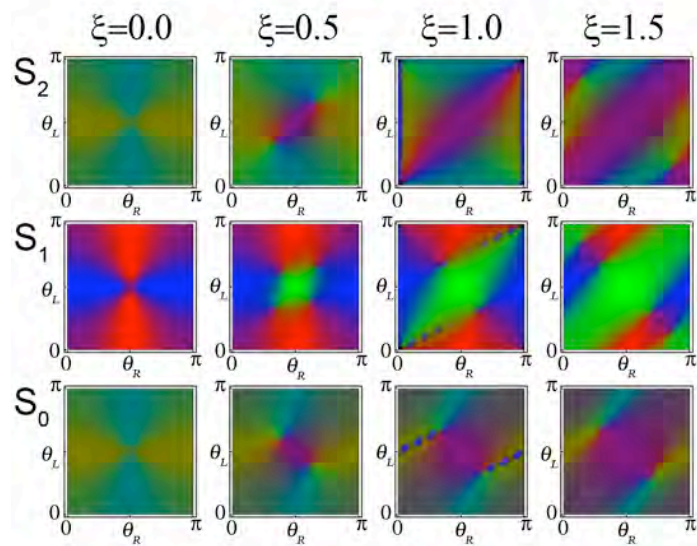

<u>The Sign of $\xi$</u>

<u>*$\xi > 0$*</u>　　　　　<u>*$\xi < 0$*</u>

*Inversion of Potential Surfaces*

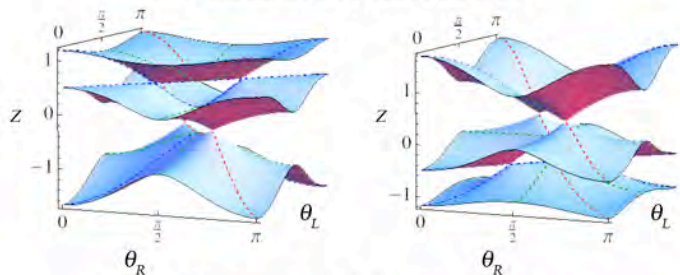

*Interchange of Intersections*

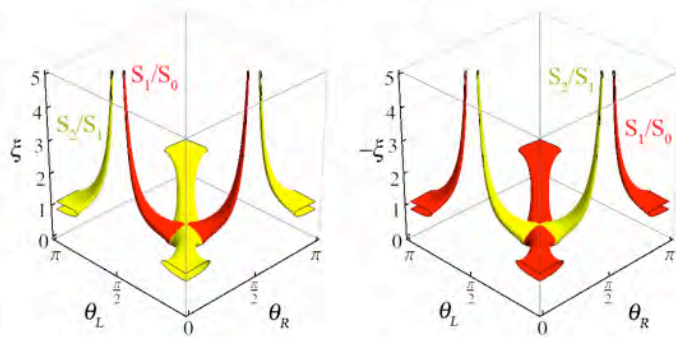